\documentclass[sigconf]{acmart} 
\AtBeginDocument{%
  \providecommand\BibTeX{{%
    \normalfont B\kern-0.5em{\scshape i\kern-0.25em b}\kern-0.8em\TeX}}}

\copyrightyear{2024}
\acmYear{2024}
\setcopyright{rightsretained}\acmConference[CHI '24]{Proceedings of the CHI Conference on Human Factors in Computing Systems}{May 11--16, 2024}{Honolulu, HI, USA}
\acmBooktitle{Proceedings of the CHI Conference on Human Factors in Computing Systems (CHI '24), May 11--16, 2024, Honolulu, HI, USA}
\acmDOI{10.1145/3613904.3642222}
\acmISBN{979-8-4007-0330-0/24/05}

\usepackage{outlines}
\usepackage{enumitem}
\setlist{  
  listparindent=\parindent,
  parsep=0pt,
}
\usepackage{todonotes}

\usepackage{graphicx}
\usepackage{caption,subcaption}

\usepackage{float}

\usepackage{xcolor}
\usepackage{array}

\usepackage{longtable}

\usepackage{makecell}

\setenumerate[1]{label=\Roman*.}
\setenumerate[2]{label=\Alph*.}
\setenumerate[3]{label=\roman*.}
\setenumerate[4]{label=\alph*.}

\usepackage{enumitem,lipsum}
\usepackage[nameinlink]{cleveref}

\usepackage{hyperref}

\usepackage{soul}

\newlist{dplist}{enumerate}{1} 
\setlist[dplist]{ref=\textbf{\unskip\arabic*},
                 label=DP\arabic*:,
                 leftmargin=*, 
                 listparindent=1.5em}
\crefname{dplisti}{dp}{dp}
\Crefname{dplisti}{\textbf{DP}}{DP}

\newlist{llist}{enumerate}{1} 
\setlist[llist]{ref=\textbf{\unskip\arabic*},
                 label=L\arabic*:,
                 leftmargin=*, 
                 listparindent=1.5em}
\crefname{llisti}{l}{l}
\Crefname{llisti}{\textbf{L}}{L}

\begin{document}


\title{TADA: Making Node-link Diagrams Accessible to Blind and Low-Vision People}


\author{Yichun Zhao}
\email{yichunzhao@uvic.ca}
\affiliation{%
  \institution{University of Victoria}
  \city{Victoria, BC}
  \country{Canada}
}

\author{Miguel A. Nacenta}
\email{nacenta@uvic.ca}
\affiliation{%
  \institution{University of Victoria}
  \city{Victoria, BC}
  \country{Canada}
}

\author{Mahadeo A. Sukhai}
\email{mahadeo.sukhai@cnib.ca}
\affiliation{%
  \institution{Canadian National Institute for the Blind}
  \country{Canada}
}

\author{Sowmya Somanath}
\email{sowmyasomanath@uvic.ca}
\affiliation{%
  \institution{University of Victoria}
  \city{Victoria, BC}
  \country{Canada}
}


\begin{abstract}

Diagrams often appear as node-link representations in contexts such as taxonomies, mind maps and networks in textbooks. Despite their pervasiveness, they present accessibility challenges for blind and low-vision people. To address this challenge, we introduce Touch-and-Audio-based Diagram Access (TADA), a tablet-based interactive system that makes diagram exploration accessible through musical tones and speech. We designed TADA informed by an interview study with 15 participants who shared their challenges and strategies with diagrams. TADA enables people to access a diagram by: i) engaging in open-ended touch-based explorations, ii) searching for nodes, iii) navigating between nodes and iv) filtering information. We evaluated TADA with 25 participants and found it useful for gaining different perspectives on diagrammatic information.

\end{abstract}





\maketitle

\section{Introduction}

Larkin and Simon define diagrammatic representations as sets of spatially distributed components with information about the relationships between components~\cite{Larkin_Simon_1987}. Diagrams often take the form of node-link representations in their different variants (e.g., boxes and arrows, dots and lines~\cite{Munzner_2014, Ware_2013, Spence_2014}), which are a common and powerful way to represent non-linear information such as people networks (e.g., social network diagrams, family trees, organizational structures), relationships between concepts (e.g., mindmaps, taxonomies) and between the constituent elements of systems (e.g., computer network diagrams).

Previous literature has claimed that these representations are challenging to access for blind and low-vision (BLV) people~\cite{Balik_Mealin_Stallmann_Rodman_2013,Petrie_Schlieder_Blenkhorn_Evans_King_O'Neill_Ioannidis_Gallagher_Crombie_Mager_etal._2002, Torres_Barwaldt_2019}. One author of this paper also has direct lived experience of this as a scientist. Lack of accessibility has a potentially detrimental effect on equity and diversity in occupations that rely on these representations, but also on the daily lives and education of BLV people. However, we are not aware of existing empirical evidence on the importance of this problem. This raises questions such as: how do people currently access information in diagrams? And, to what extent do they face challenges and find workarounds?

We carried out a formative interview study with 15 people to answer these questions and to inform possible solutions. We found that lack of accessibility of diagrammatic representation is a frequent issue affecting BLV people, that their circumstances and motivations to access these are varied and multi-faceted and that existing solutions such as alt-text, asking other people and finding multiple sources for the same information are either insufficient or costly in terms of time, effort and social capital.

As a logical next step, we set out to integrate what we learned and existing experiences from literature (e.g. TeDUB~\cite{King_Blenkhorn_Crombie_Dijkstra_Evans_Wood_2004,Horstmann_Lorenz_Watkowski_Ioannidis_Herzog_King_Evans_Hagen_Schlieder_Burn_etal._2004,Petrie_Schlieder_Blenkhorn_Evans_King_O'Neill_Ioannidis_Gallagher_Crombie_Mager_etal._2002,Petrie_King_Burn_Pavan_2006}, A11yBoard~\cite{Zhang_Wobbrock_2023} and GeoTablet~\cite{Simonnet_2012}) to design a new interface that further advances the state of the art. This tool goes beyond existing solutions by: a) preserving spatial relationships from the diagram; b) enabling rich and multi-level access to the information in varied ways; and, c) retaining affordability. 
TADA runs on inexpensive tablets and integrates touch and speech input and audio output. It provides a range of interaction techniques on node-link representations to access overviews, navigation, information detail, filtering and finding.

    


We evaluated the atomic elements of the design with 25 people, which confirmed the practicality of our approach and offered ways to fine-tune the design. Our main contributions are:
\begin{itemize}[topsep=6pt]
    \item The results of an interview study on the importance of accessing diagrammatic information for BLV people.
    \item The empirically validated design of an interface that allows BLV people to access node-link diagrams.
\end{itemize}

\section{Related Work}
\label{sec:related_work}

This section surveys existing systems and techniques that make visual content accessible to BLV people, with a focus on diagrams. We use the term \emph{diagram} to refer to graphics that represent objects and their relationships, such as node-link diagrams, flowcharts and UML diagrams. Diagrams can be stored as \emph{graphs}. A graph is a mathematical abstraction and an abstract data structure that can be represented in several ways, one popular way being \emph{node-link diagrams}. We use the term \emph{chart} for statistical visualizations such as line and bar charts.

For input/output, we use \emph{tactile} to refer to systems that use the sense of touch for feedback, such as raised-line overlays or pin arrays. We use \emph{haptic} when there is force feedback, including through vibrotactile actuators. \emph{Touch} refers to input from the location of bodily contacts with a surface (usually fingers on screens), even when it does not provide dynamic feedback through those contacts. \emph{Proprioception} refers to people's awareness of the position of their own limbs (e.g., sensing a finger's position relative to another finger).

\subsection{Non-spatial Systems}
Screen readers (e.g., JAWS, NVDA) ~\cite{AFB-screen-readers} read aloud \emph{alternative text} and \emph{image descriptions} to relay details about the visual content to BLV people. There are well-established guidelines for the generation of alt-text~\cite{Guidelines-Image-Descriptions, Image-Description-Guidelines, Jung_Mehta_Kulkarni_Zhao_Kim_2022, Kim_Kim_Kim_2023, Edwards_Gilbert_Blank_Branham_2023, Initiative-WAI}, and we see a recent push to generate descriptions automatically~\cite{Cross_Cetinkaya_Dogan_2020, Hoppe_Morris_Ewerth_2021, OpenAI_2023, Hearst_2023, Choi_Jung_Park_Choo_Elmqvist_2019}. Guidelines exist to improve accessibility for diagrams~\cite{Torres_Barwaldt_2019} as well as machine learning systems that automatically convert images of diagrams into machine representations~\cite{Kembhavi_Salvato_Kolve_Seo_Hajishirzi_Farhadi_2016, Kim_Yoo_Kim_Lee_Kwak_2018}. The quality of descriptions can make a difference in their usefulness, however, regardless of quality, alt-text often provides a description that is limited to a single purpose (e.g.,  a general description that will not suffice for someone looking for detail and a detailed description which is tedious for someone with a passing interest). 
Kim et al.~\cite{Kim_QA_2023} identify this and other limitations of these approaches and suggest an approach (simulated through \emph{Wizard of Oz} techniques) where instead of a fixed description, users can query figures through natural language questions.

Further evolution of descriptions includes systems that provide hierarchical arrangements of information, which come closer to a continuous spatial approach. The user can move up, down or sideways in the hierarchy, usually with keyboard direction keys. This has been applied to UML and technical diagrams~\cite{King_Blenkhorn_Crombie_Dijkstra_Evans_Wood_2004, Petrie_Schlieder_Blenkhorn_Evans_King_O'Neill_Ioannidis_Gallagher_Crombie_Mager_etal._2002, Horstmann_Lorenz_Watkowski_Ioannidis_Herzog_King_Evans_Hagen_Schlieder_Burn_etal._2004}, display floor plans~\cite{Petrie_King_Burn_Pavan_2006} and chemical molecule representations~\cite{Brown_Pettifer_Stevens_2003}. 

Another example of a system that represents diagrams without preserving spatial relationships is Kevin CASE, which uses a tactile overlay to interact with a table realignment of data flow diagrams for software engineers~\cite{Blenkhorn_Evans_1998}. 

Our research differs fundamentally from these approaches because we support spatial perception of the continuous arrangement of data. We believe it leverages spatial memory and conveys important information (see Section~\ref{sec:goals} for further justification).

\subsection{Chart Readers}
A group of related systems addresses chart accessibility (e.g., line or bar charts). An initial set of such systems are image description generators specifically for charts~\cite{Cross_Cetinkaya_Dogan_2020, Hoppe_Morris_Ewerth_2021, OpenAI_2023, Hearst_2023, Choi_Jung_Park_Choo_Elmqvist_2019}. More interactive systems provide translations of data to sound and speech that are navigable and therefore better adaptable to the user's particular needs. These systems differ in how the data is rendered. For example, SeeChart~\cite{Alam_Islam_Hoque_2023} provides a sophisticated way to navigate D3 charts\footnote{D3: \url{https://d3js.org/}} via keyboard, with speech output only and VoxLens has a similar approach but adds speech as input modality~\cite{Sharif_Wang_Muongchan_Reinecke_Wobbrock_2022}. Infosonics~\cite{Holloway_Goncu_Ilsar_Butler_Marriott_2022} provides both audio sonification and speech controlled by keyboard. 
Other systems such as Chart Reader~\cite{Thompson_Martinez_Sarikaya_Cutrell_Lee_2023} and Zong et al.'s work~\cite{Zong_Lee_Lundgard_Jang_Hajas_Satyanarayan_2022} support accessible interactions of charts with screen readers. 
TactualPlot~\cite{Chundury_Reyazuddin_Jordan_Lazar_Elmqvist_2023} facilitates the exploration of scatter-plots through touch and audio. Finally, Slide-Tone and Tilt-Tone use a novel haptic output device to enhance the haptic and sound perception of line charts~\cite{Fan_Siu_Law_Zhen_O'Modhrain_Follmer_2022}. 

Our work differs from chart readers in the type of data represented; we focus on diagrams (formalizable as relationship graphs), whereas most work in charts focuses on the representation of specific types of visualizations such as line and bar charts, which require different interactions and rendering.

\subsection{Map Accessibility Systems}
Many systems have been proposed to represent maps non-visually. One group provides static tactile feedback (a printed raised-line overlay) with finger tracking, which allows the system to react differently (e.g., with sounds) to different finger positions. NOMAD~\cite{parkes1988nomad}, Mappie~\cite{Brock_Truillet_Oriola_Jouffrais_2014, Brule_Bailly_Brock_Valentin_Denis_Jouffrais_2016} and the tactile graphics helper~\cite{Fusco_Morash_2015} all use variants of this input/output combination. In some cases, instead of fingers, the system senses tangible objects above the surface~\cite{Ledo_Nacenta_Marquardt_Boring_Greenberg_2012,Marquardt_Nacenta_Young_Carpendale_Greenberg_Sharlin_2009,Ducasse_2016}. The audio output can be speech (Brock et al.'s system names double-tapped geographical objects~\cite{Brock_Jouffrais_2015}), or iconic sounds (Albouys-Perrois et al.'s system~\cite{Albouys_Perrois_Laviole_Briant_Brock_2018} beeps to indicate correct location of objects). 
Replacing static printed overlays with array pins overcomes the fundamental limitation of static displays (which we also consider necessary to avoid in our design). Examples include Audio-Haptic browser~\cite{Zeng_Weber_2010}, ATMap~\cite{Zeng_Weber_2012}, as well as Shimada et al.'s~\cite{Shimada_Murase_Yamamoto_Uchida_Shimojo_Shimizu_2010}, Schmitz et al.'s~\cite{Schmitz_Ertl_2012}, Ivanchev et al.'s~\cite{Ivanchev_Zinke_Lucke_2014}, and Holloway et al.'s~\cite{Holloway_Ananthanarayan_Butler_DeSilva_Ellis_Goncu_Stephens_Marriott_2022} devices. Earlier systems are summarized by Vidal et al.~\cite{Vidal-Verdu_Hafez_2007}. 
A key drawback of these is their specialized hardware, which is still relatively rare, expensive and low-resolution.


Finally, some systems depict maps without relying on haptic means, which makes them more affordable~\cite{Ducasse_Brock_Jouffrais_2018}. Many combine computer-generated audio with vibrotactile feedback. Geotablet~\cite{Simonnet_2012} plays a range of speech, iconic and environmental sounds when touched and TIKISI~\cite{Bahram_2013} offers a range of interesting interactions such as zooming, selection and overviews. TouchOver Map~\cite{Poppinga_2011}, Timbremap~\cite{Su_Rosenzweig_Goel_De_Lara_Truong_2010} and Open touch/sound maps~\cite{Kaklanis_Votis_Tzovaras_2013} have similar approaches. 

We took inspiration from many of these designs. For example, Yairi et al.'s system~\cite{Yairi_Takano_Shino_Kamata_2008} guides fingers from one node to another using pitch-modulated sound.
However, map representation poses fundamentally different challenges to diagram representation: in maps, there is more emphasis on the shape and location of objects whereas in diagrams the representation of explicit relationships between different objects is key (in maps, these relationships are most often implicit). Consequently, in maps, egocentric navigation is often a key task, which is, arguably, not as relevant for diagrams. 

\subsection{Images and Raster Renderers}
Some researchers have attempted non-visual rendering of generic images (e.g., photographs) to make them accessible through non-visual channels. Haptic renderings (e.g., pin arrays) of images might be too low resolution to effectively allow recognition of elements in raster representations, but with pan, zoom and animation~\cite{Holloway_Ananthanarayan_Butler_DeSilva_Ellis_Goncu_Stephens_Marriott_2022} or clever pre-processing (e.g., filtering or simplification~\cite{Rastogi_Pawluk_2013}) the issues can be ameliorated. Some have also tried audio renderings, for example, by mapping brightness to sound in a sweep of an image~\cite{NASA-2020,Bragard_Pellegrini_Pinquier_2015}, by sonifying the brightness of a raster image pixel by pixel depending on touch position~\cite{Uno_Suzuki_Watanabe_Matsumoto_Wang_2018}, or by translating colours and edges of objects into the sound of different instruments~\cite{Banf_Mikalay_Watzke_Blanz_2016}. This group of systems informs audio mapping in our work, but our approach differs fundamentally because our representation level is objects rather than pixels. 

\subsection{Vector- or Object-based Spatial Systems}

Maćkowski et al. presented a system based on a raised-line overlay on top of a tablet to represent SVG graphics that users can query at different levels of detail by tapping~\cite{Spinczyk_2023}. \emph{Ad hoc} overlay creation can be tedious, which can benefit from a toolchain as proposed by He et al.~\cite{He_Wan_Findlater_Froehlich_2017}. 
Linespace~\cite{Swaminathan_Roumen_Kovacs_Stangl_Mueller_Baudisch_2016} and PantoGuide~\cite{Chase_Siu_Boadi-Agyemang_Kim_Gonzalez_Follmer_2020} are hardware-intensive tactile technologies. Linespace uses 3D printing to create raised line contours on a large surface. PantoGuide enhances static tactile graphics like charts with audio and a hand-mounted system, providing extra tactile information, such as the direction of a search target based on finger location.
Also based on a glove, GraVVITAS~\cite{Goncu_Marriott_2011} provides multi-finger vibrotactile feedback on top of a surface based on hand position.

Moving away from systems that require printing or use specialized hardware, Giudice et al. proposed an alternative to access audio descriptions of objects where, instead of a raised line, the high-end tablet vibrates in different ways when the finger traverses different types of graphical elements~\cite{Giudice_Palani_Brenner_Kramer_2012}. Alty and Rigas~\cite{Alty_Rigas_2005} carried out a relevant study where they investigated encoding the position of objects in the plane through audio timbre and pitch (not integrated into a system).

Two recent relevant systems that use only touch position as input and audio as output are A11yBoard~\cite{Zhang_Wobbrock_2022, Zhang_Wobbrock_2023} and ImageAssist~\cite{Nair_Zhu_Smith_2023}. These relate to our own approach in that they provide more sophisticated sets of interaction techniques to access the data (segmented and semantically annotated pictures and artwork for ImageAssist and artboards for A11yBoard). The main difference between these work and ours is that the interaction techniques necessarily differ because of the different types of target graphics.

\subsection{Systems for Node-link and Diagram Display}
A seminal example of a system for representing node-link diagrams is Audiograf, which provides speech descriptions of objects and other auditory cues under the finger of the user in a touch input device~\cite{Kennel_1996}. There are three systems that enable interaction with node-link diagrams for BLV people and go beyond the simple output of Audiograf. TeDUB~\cite{King_Blenkhorn_Crombie_Dijkstra_Evans_Wood_2004,Horstmann_Lorenz_Watkowski_Ioannidis_Herzog_King_Evans_Hagen_Schlieder_Burn_etal._2004,Petrie_Schlieder_Blenkhorn_Evans_King_O'Neill_Ioannidis_Gallagher_Crombie_Mager_etal._2002,Petrie_King_Burn_Pavan_2006} and GSK~\cite{Balik_Mealin_Stallmann_Rodman_2013,Balik_Mealin_Stallmann_Rodman_Glatz_Sigler_2014} share interaction through keyboard or joystick and therefore support navigation of diagrams that are relative (i.e., the current node's spatial position has to be updated mentally by the user as they traverse the diagram). This is fundamentally different from how our system and PLUMB~\cite{Cohen_Yu_Meacham_Skaff_2005,Calder_Cohen_Lanzoni_Xu_2006,Cohen_Haven_Lanzoni_Meacham_Skaff_Wissell_2006,10.1145/1124706.1121428} operate, where the input device is absolute (stylus for PLUMB and fingers on a tablet for TADA), which makes the current location of exploration continuously accessible to the user through proprioception of their own limbs's position with respect to the device. However, TeDUB allows annotation, and GSK and PLUMB allow creation and editing of links, which we do not consider at this stage. PLUMB supports, however, relatively few interactions to perceive the diagram (it supports only object rendering and link following).  

\subsection{Interaction Techniques}
A few researchers have studied and compared different ways to interact in audio or vibrotactile output touch location-based systems for specific tasks. For example, Kane et al. investigate invocation of commands on mobile phones for blind users~\cite{Kane_Bigham_Wobbrock_2008}, Tennison and Gorlewicz~\cite{Tennison_Gorlewicz_2019} compare techniques and strategies to perceive and follow lines, Ramôa et al.~\cite{Schmidt_2022} and Kane et al.~\cite{Kane_Morris_Perkins_Wigdor_Ladner_Wobbrock_2011} compare ways to direct users to elements on the 2D space, with the latter also covering browsing regions and retrieving localized detail. We integrated this knowledge into the design of our system.

\section{Formative Interviews: Challenges with and Strategies for Diagram Accessibility}\label{sec:formative_study}

To understand how individuals currently access diagrams (node-link representations), including their challenges and strategies for access, we conducted remote semi-structured interviews with 15 individuals. 

\subsection{Participants}
Fifteen adults (8 men and 7 women) ages 18 to 69 years (mean=47.8, median=50, SD=14.17) participated in our 50-90 minute-long interview study. Nine participants (P1-4, P6, P8, P11, P13, P15) are totally blind. Six of them 
were born totally blind and three 
lost their vision in adulthood. Six other participants (P5, P7, P9, P10, P12, P14) are legally blind or have low vision, with five of them 
having congenital or early-onset vision loss, and one 
experiencing recent vision loss. 
Appendix \ref{appendix:formativeStudy:ParticipantDetails} includes other participant details. 
Our inclusion criteria required: 1) self-identification as legally blind; 2) reliance on the auditory channel or a combination of auditory and other sensory channels for accessing digital information; and 3) utilization of screen readers, braille-based systems, self-devised solutions, or alternative methods for digital information access. All participants encountered diagrams in personal, professional, or educational contexts, either daily or sporadically. 
Participants were asked to think about three examples of diagrams they had encountered in their lives prior to the study sessions and bring digital examples. 
Participants received CAD\$30 for their time. The study was pre-approved by the local Research Ethics Board (REB).

\subsection{Procedure} 
The experimenter collected participant information on age, gender, background and vision loss. Then, participants described their encounters with diagrams and, using those as references, answered questions regarding information accessibility. The questions focused on the diagram, the context of the encounter, the importance of the information, motive(s) for accessing the diagram, tools used (if any), the types of tasks when using the diagram, the challenges encountered and their workarounds. The detailed study procedure is in Appendix \ref{appendix:formativeStudy:InterviewDetails}.

\subsection{Data Collection and Analysis}
We analyzed 16 hours of interview data using the thematic analysis method~\cite{Adams_Lunt_Cairns_2008, boyatzis1998transforming} and iteratively coded the interview transcripts. First, one researcher open-coded two transcripts to bootstrap the codebook. 
With this initial set of codes, we collaboratively discussed and refined the codebook as we coded the remaining transcripts. This resulted in a total of 124 codes grouped into the themes we discuss below. 

\subsection{Findings}
\label{sec:formative:findings}

The most basic findings of our study validate the importance of diagrams for our participants. They showed us a multiplicity of diagrams and how they interact and struggle with them, such as textbook diagrams [P2-6, P9, P11, P14], networks and flowcharts [P1, P5, P10, P12, P14] and instructional diagrams [P2, P5, P11]. Participants also highlighted the personal and professional significance of many diagrams, arguing that diagrams enhance existing information [P1, P14], facilitate concept learning [P2-P6, P8, P12] and reflection [P12], are indispensable in certain professions [P1, P2, P7, P11], support recreational and personal activities [P8, P10, P11, P13] and enable collaborations with sighted people [P7, P12, P15].


Their challenges with diagrams were varied. They often missed their existence [P1, P4, P7-8, P12, P15], had unequal access [P1-2, P4-5, P8, P10-15],
lacked control over the required level of detail [P1-5, P7-8, P10-12, P14], experienced unfair cognitive burdens such as having to memorize large chunks of information [P1-2, P4-5, P8-9, P11, P13-15] and incurred additional social costs such as those derived from bothering sighted colleagues or acquaintances [P1-3, P8-11, P13-14]. To overcome these challenges, they employ strategies such as asking sighted people for help [P1-13, P15], using OCR/CV tools [P4-5, P7, P9, P15] and finding multiple sources for the same information [P2-4, P6, P9-15]. 

Participants' comments were rich and insightful. During analysis, we realized that data themes aligned well with a linear structure of the participants' increasing ability to access information that we call the \emph{ladder of diagram access}, which we use to summarize and organize our description of the insights in the data. 

\subsubsection{Ladder of Diagram Access}
\label{sec:formative:ladder}

We describe five levels of diagram access, from no access at all (level 1) to comprehensive (level 5). Within each level, we elaborate on aspects related to 1) individual differences in terms of abilities, motivations or tasks they wanted to accomplish, 2) types of diagram properties they attempted to access, and 3) external factors that influenced their process of accessing a diagram.

\begin{enumerate}[leftmargin=16pt]
\item[L1:] \label{l1} \emph{Not Knowing the Existence of a Diagram}:
In our study, 7 of 15 participants explicitly told us about experiencing a situation wherein they did not know if a diagram existed. For instance, P12 referred to a document with a diagram and showed us how, when accessed via a screen reader, it skipped the diagram completely: \textit{``It tells us everything on the other [text] pages, and then you hit the page with the diagram, and it stops."} Similarly, P15 commented \textit{“I honestly wouldn’t be able to tell you if there was a diagram there [on a webpage] or not.”} We classified such instances of interaction as belonging to level 1, which corresponds to the bottom rung on our ladder and refers to the most basic level of information access wherein people have to know the existence of an information representation. 

\item[L2:] \label{l2} \emph{Knowing the Existence of a Diagram}:
In our study, 2 participants explicitly described situations wherein an individual is made aware of the presence of some kind of embedded object in a document or website, but no further information is available. For example, during the interview, P9 showcased how a screen reader only informed them about a ``graphic'' and commented \textit{“[it] means nothing to me”} and \textit{“there's basically nothing to give [me] an idea [of the diagram]”}. Similarly, P14 showed us a diagram in a PowerPoint slide which read “embedded object” by the screen reader, signalling that there might be information, but the placeholder failed to inform P14 if the embedded object was a diagram, leaving P14 confused and \textit{“[having] no idea what that [meant].” }

Without such access to the actual diagrammatic information, people cannot decide if the diagram is relevant to their needs and if they should access or ignore the diagram. P10 commented \textit{“[I’m] constantly making the assessment [of] how worthwhile [a diagram is] for me to understand”} and four other participants confirmed this, but they cannot do so with just a placeholder. 

We refer to such interactions as level 2 information access, wherein people know there is a placeholder but have too little information to determine whether the information in the visual element is important. 

\item[L3:] \label{l3} \emph{Single Static Perspective of a Diagram}:
Five participants in our study described examples wherein they were aware that there was a diagram and were able to learn more about it. This type of learning usually involves a static feature like alt-text. For example, P7 commented: \textit{“If the human-generated alternative text is done well, then you can gain a very good picture of the diagram.”} P7 and P9 used new computer vision technology to learn more about the diagrams. More commonly, a majority (14 of 15 participants) resorted to asking sighted people but found that this workaround came with challenges (discussed in level 4). 

Despite the benefits of static descriptions, 14 participants also voiced concerns over not having enough control over them, which limited how much they could gain. For example, P12 explained that an inherent limitation of using screen readers is that one does not have control over the types of details one wants to access: \textit{“[...] Sometimes, you just want to skim something, and you can't. You can't do the `cheaty' short[cut] method of looking at the diagrams [visually] and inferring the information.” } Another limitation of text-based descriptions such as alt-text, image descriptions and Optical Character Recognition (OCR) readouts was raised by P5 and P14, who mentioned how the linearity of these aids forces people to piece together different parts of the information on their own, which can be cognitively cumbersome. For example, P14 demonstrated accessing a node-link diagram about Belsky’s (1984) Process Model of the Determinants of Parenting found in a textbook through OCR. P14 described how, while they were learning about the details in the nodes through the textual descriptions, they missed the overall structure and when focusing on the structure, they could not remember the details. 

Lastly, P3 and P10 complained about how the information made available is not necessarily the information one needs. P10 said, \textit{“In some instances, I could see individuals who [...] want to understand the design or want to understand the layout, and they just don't want the [provided] statistical information [...] They want to understand the lines, the flows and that sort of information. So it's a big challenge.”} P3 repeatedly accessed the same diagram to enhance understanding but found the process frustrating: \textit{“I would try until the bitter end until it looks like I can't handle this anymore, like I’m frustrated.”} Another example of the mismatch between a static description and the user purpose is P3's comment: \textit{``I just scroll past and don't really think about it again, unless its information that's important to me, and then I get frustrated that either it's not all there, or it's not written in a way that [...] immediately makes sense."}

Six participants in our study described how they sometimes needed to find information through complementary sources. For instance, P2 mentioned finding other people's written journals when unable to access a diagram of hiking routes. P9 searched for alternative diagrams with text descriptions on search engines once they knew a bit more about the original diagram. However, this workaround is time-consuming and challenging. For example, P3 complained: \textit{``It can [be time-consuming and] also frustrating because [the information should be] right there and everybody else has it.''} 

We classify static descriptions and OCR output as belonging to level 3, wherein an individual knows that a diagram exists and is able to acquire at least one perspective from the diagram through these aids. Using such information, individuals are also able to complete some tasks. However, the lack of control over what information is available makes gaining a deeper understanding challenging and frustrating. 

\item[L4:] \label{l4} \emph{Multiple Perspectives of a Diagram}:
Eleven participants in our study described scenarios when they attempted to acquire additional perspectives to complement static descriptions (such as those belonging to level 3). For example, P15 often wants a global perspective first and only after a look at certain details: \textit{“You have to tell the basic first and then you tell the details of it.”} P7 depicted how he achieves different perspectives by reorganizing the information in their own head: \textit{``Sometimes you have to come up with new ways of representing things, right? So that you know this relates to this, but in an alternative way it relates to that as well."}

Participants often reported scenarios wherein they reached out to sighted people (through an app, or in person) to acquire missing perspectives (n=14). For example, P5 uses “Be My Eyes”~\cite{BeMyEyes}, which allows volunteers to access the mobile phone camera to help them. This way they can ask targeted questions to gain more understanding. Similarly, P3 uses Twitter: \textit{“I would probably just tweet the person and [...] ask a specific question about [the diagram].”} and P8 mentioned having a direct dialogue with architects to clarify the exact structure of a building.

While such external help was meaningful, 9 participants also commented on the challenges of asking others. For example, the volunteer might not have the necessary expertise, patience or availability to explain the diagram well. P8 and P10 also explicitly highlighted the social costs. P8 told us how \textit{``you can sometimes drive people crazy when you really want to understand something, and some people don't have the ability to explain it to you easily. People get frustrated because they want you to understand their explanation, and if you're not getting it, then there's a frustration level.” } 

We associate these interactions with level 4. Individuals are able to learn more about the diagram than in levels 1-3 by acquiring multiple perspectives of the information and querying. While the mechanisms for querying are not always convenient or sufficient, our participants find that reaching this point in their information access journey is helpful nevertheless.   

\item[L5:] \label{l5} \emph{Effective and Comprehensive Access to Diagrams}:
As per the subsections above, BLV people are able, through existing technologies or workarounds, to access information in diagrams to accomplish virtually any required task. 

However, the ability to access information still does not guarantee equity of information access for BLV people because current aids or strategies, although enabling, are costly in ways that can negate the benefits which motivate representing information as diagrams in the first place. For example, we know that visuals can reduce cognitive load for sighted individuals by, for example, allowing them to rely on spatial memory references instead of having to remember details~\cite{ballard_hayhoe_pook_rao_1997} and the general advantages of diagrammatic representations are well known (e.g.,~\cite{Larkin_Simon_1987,scaife1996external,BINKS2022102851}). Many of these advantages likely disappear if access is mediated in slow or cognitively expensive ways. P2 put this eloquently: \textit{``What is accessibility? Is accessibility just putting something out there? Or is it being able to give the person the same access, so that they are able to generate, have the same opportunity to get an equal level of understanding as another person?''}. P14 reflected that \textit{``a person with sight really could like, boom, they could kind of answer the simple question really quite quickly, just like looking visually at the [diagram ...] For me, it took more time and effort.''}

The final level of the ladder is therefore defined by the efficiency and effectiveness with which people can access the information, not just by whether the information itself is accessible assuming almost unlimited supplies of cognitive abilities, time and effort. The system that we describe in the following sections intends to bring people \emph{closer} to this level of accessibility, but we can only aspire to, not claim success in, making diagrams accessible at this level.

\end{enumerate}

\section{Design Goals and Principles}\label{sec:goals}

We set out to design and implement a tool that accomplishes two main overarching goals: \textbf{G1}: to help people access diagrammatic information in richer and more interactive ways (consistent with higher rungs of the ladder of diagram access---Section~\ref{sec:formative:ladder}) and \textbf{G2}: to reduce common access barriers to this information, such as steep learning curves and expensive or unavailable hardware.

To achieve these goals we selected a set of design principles that guide our design. We chose the design principles to encode what we learned from the formative study (Section~\ref{sec:formative_study}) and to apply existing knowledge from previous research (e.g.,~\cite{Tennison_Gorlewicz_2019,Schmidt_2022, Hansen_BaltaxeAdmony_Kurniawan_Forbes_2019}).
We applied the overarching goals and the design principles in all stages of the design process, from key foundational choices at the top level (e.g., which modalities to support), to the detail of interaction techniques. The design principles and goals are referenced in the next section to justify design decisions and features (e.g., \textbf{G1} or \textbf{DP3}). 

\begin{enumerate}[leftmargin=22pt]
\item[DP1:] \label{dp:spatiality} \emph{Rely on Spatiality}: Diagrams encode information in the spatial relations between the display objects as well as between objects and the diagram's frame. For example, objects proximate in visual representations are interpreted as more related to each other than distant objects~\cite{Mayer_2002, williams2015non, Larkin_Simon_1987}, and diagram designers often use layout to convey intended reading order, importance and other attributes (e.g., in an organization chart the CEO is at the top). We also know that viewers encode elements spatially in memory for agile retrieval while engaging with the visual representation~\cite{ballard_hayhoe_pook_rao_1997}, hence the importance of the \emph{mental map} in studies of network visualization~\cite{Archambault_Purchase_2013, Montello_2015, Brock_thesis}. People without vision also rely on spatial object layout, they just do not access it visually, relying instead on other modalities~\cite{Chundury_Patnaik_Reyazuddin_Tang_Lazar_Elmqvist_2022} such as proprioception~\cite{Yamamoto_Shelton_2005}, touch~\cite{Ducasse_Brock_Jouffrais_2018} or echolocation~\cite{Andrade_Waycott_Baker_Vetere_2021,thaler2016echolocation} to build their spatial understanding.
In terms of the ladder, spatiality supports the transition from \textbf{L3} to \textbf{L4} because it enables multiple alternative points of access, akin to visual gaze for visual interfaces. 
Additionally, preserving spatiality benefits both sighted and BLV people, enabling communication through spatial deixis. 

One of our key assumptions is that spatial layout and memory can be effectively leveraged by BLV users. This is consistent with Chundury et al.'s interview study with mobility experts~\cite{Chundury_Patnaik_Reyazuddin_Tang_Lazar_Elmqvist_2022}, which highlights the importance of spatial awareness to understand data visualizations. 

\item[DP2:] \label{dp:information_types} \emph{Make Element Categories Perceptually Different}: Visual diagrams map certain types of elements (e.g., vertices, links and names in a network structure) to specific visual elements (e.g., red square boxes, arrows between boxes and text). An important, but often implicit, characteristic of these mappings is that elements are easily distinguishable from each other~\cite{Kosslyn_1989}. For example, nodes are immediately distinguishable from other elements such as labels or connections. Unlike in visual diagrams, where a long history of conventional mappings exists, other modalities require careful selection of representations in the output space (in our case, sounds) that are fast and clearly distinguishable from each other~\cite{Chundury_Patnaik_Reyazuddin_Tang_Lazar_Elmqvist_2022}. This works towards achieving quicker and more efficient understanding [\textbf{L5}].

\item[DP3:] \label{dp:levels_of_information} \emph{Support Multiple Levels of Information Access}: Our interviews showed that people have diverse varying circumstances and motives to access diagrams. For example, sometimes one wants to gauge the amount of information in a diagram (e.g., to decide whether to study it in more detail later). In some cases, the structure of the diagram is most important, whereas sometimes the specific details are crucial. To support people's varied motives and circumstances, the design needs to implement multiple forms of access to information [\textbf{L4}], including summaries, overviews, finding and navigation, thereby allowing BLV people to choose the level of detail and granularity. 
As we have seen in Section~\ref{sec:related_work}, most existing systems do not provide comprehensive coverage for the variety of actions necessary in a wide variety of scenarios. 

\item[DP4:] \label{dp:multiple_interactions} \emph{Complementary and Simultaneous Interactions}: As a consequence of \textbf{DP3}, it follows that users should be able to choose the way they engage with diagrams depending on their purpose and circumstances, providing flexibility and choice in how they interact with diagrams [\textbf{L4}] as well as efficiency [\textbf{L5}]. This suggests an approach with multiple interaction techniques, perhaps some of which to achieve the same goal but from a different starting point or with different levels of precision. The interaction techniques could also be complementary or simultaneous; we know that users of tactile maps often use multi-handed operations with one finger or hand being used to keep memory~\cite{Ducasse_Brock_Jouffrais_2018, Tennison_Gorlewicz_2019} or as a ``bookmark''. Therefore we seek to design techniques that seamlessly take place simultaneously or compose and succeed each other. 

\end{enumerate}

\section{TADA's Design} 

Here we describe TADA's design, starting with a general overview followed by a description of its interaction techniques.

\subsection{System Overview}

TADA is currently implemented as a multi-platform application with Unity~\cite{Unity_Technologies} and ChucK~\cite{Atherton_Wang, Wang_Cook_2003} (See Appendix~\ref{appendix:implementationDetails}). It is designed for tablet devices equipped with an inbuilt microphone and speaker. TADA's input is touch (locations of fingers touching the tablet's surface) and audio (speech commands), and its main output are sounds and speech synthesized in real time. The screen also shows a simple representation of the diagram for people with low or residual vision and for collaboration with sighted individuals (see~\cite{Fan_Glazko_Follmer_2022}). The choice of platform and modalities is consistent with \textbf{G2}: it requires no specialized hardware and it can run on cheap tablets or even phones (although smaller screens will reduce leverage of~\textbf{DP1}---see Subsection~\ref{sec:discussion:TADADesign}). TADA does not rely on vibrotactile input because most tablets do not offer it; however, the current design is compatible with adding this modality.

TADA is designed to access node-link diagrams; therefore the main objects that it represents are nodes, their connecting links, and the information contained in both (labels and attributes).

\subsection{Design Process}

We followed an iterative user-centered approach to which each author contributed, including one of the authors who is a BLV scientist and diagram user. The process gradually and iteratively transitioned from sketching, low-fidelity prototyping, to high-fidelity prototyping following Buxton's design process model~\cite{buxton2010sketching} over the course of 36 design sessions each ranging from 30 minutes to one hour. We devoted two one-hour sessions specifically for testing by our BLV author. In later stages of the design process the focus shifted to the refinement of interaction techniques to work well in conjunction with each other and adjusting threshold values to reasonable defaults (e.g., size of objects and touch drift tolerances). We regularly tested the design with different types of diagrams (e.g., networks and textbook diagrams), to accommodate a variety of node-link representations. We built a randomized diagram generator to help evaluate the robustness, adaptability and scalability of the prototypes.

\subsubsection{Sound Design}
The main output is audio of two kinds: abstract and speech. Following \textbf{DP2}, we map speech and abstract sound (which are easily distinguishable from each other) to different kinds of information: speech renders information contained in nodes and links (node and link names, attributes) and non-speech sounds (tones and musical instrument notes) signal the main objects and their structure. Within the non-speech sound category, we dedicated significant time to finding easily distinguishable sounds. 
We tested mappings from previous sonification literature~\cite{Hansen_BaltaxeAdmony_Kurniawan_Forbes_2019, de2007toward}, environmental sounds, and a variety of simple waveforms (e.g., sine, triangle, saw-tooth) but we settled on utilizing musical instruments to create different timbres, balancing distinguishability, pleasantness and appropriateness. 

A node produces a french-horn sound, with higher-pitch representing nodes with more connections. A link produces a plucked guitar sound, and a higher-pitched sound means it is shorter than other links (consistent with how longer strings vibrate at lower frequencies). A sustained french-horn or guitar sound indicates dwelling on a node or link respectively; A discrete sound indicates sweeping over a node or link. A continuous 150Hz pure tone (neither horn nor string) indicates the proximity to a diagram element. 

\subsection{Interaction Techniques}

\begin{figure*}[htp]
    \centering
    
    \begin{subfigure}[b]{0.39\textwidth}
        \centering
        \includegraphics[width=1\textwidth]{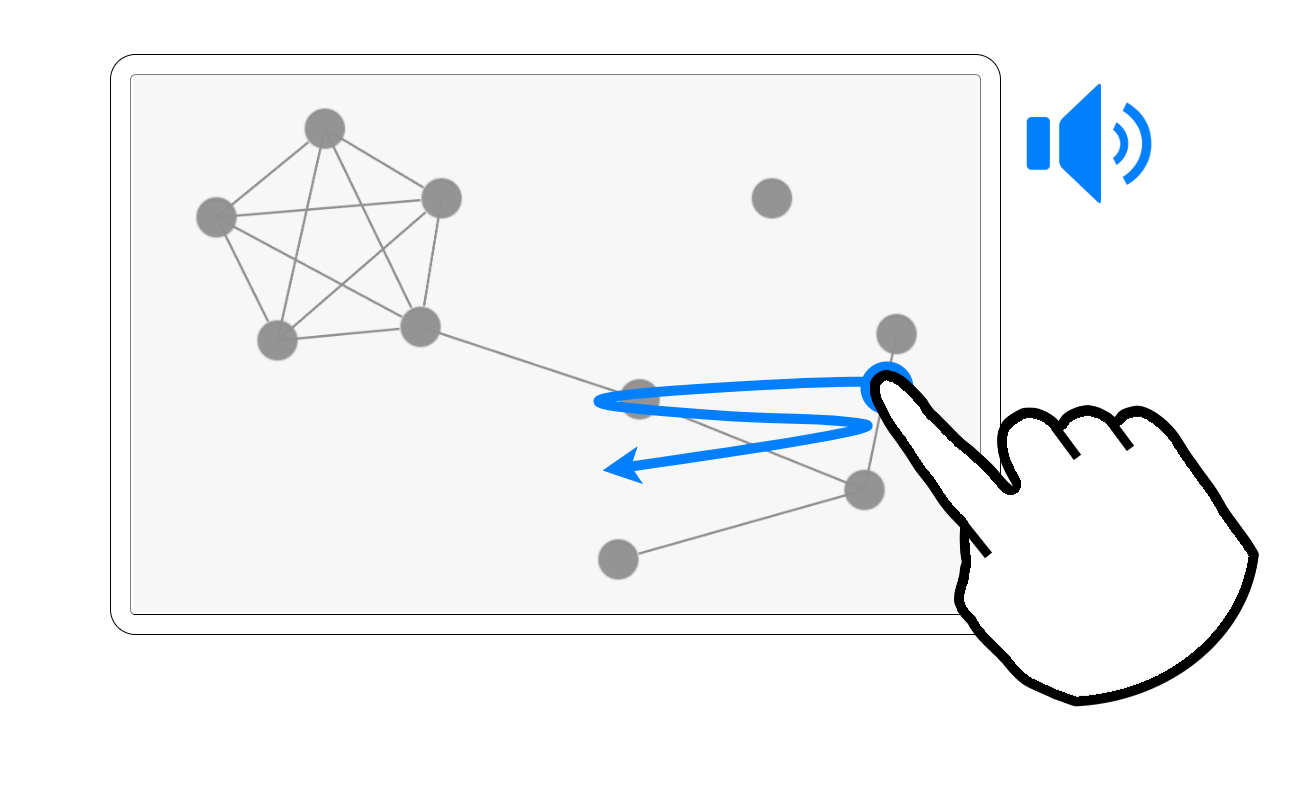} 
        \subcaption{Single-finger Sweep.}
        \label{fig:single_finger_sweep}
    \end{subfigure}
    \begin{subfigure}[b]{0.39\textwidth}
        \centering
        \includegraphics[width=1\textwidth]{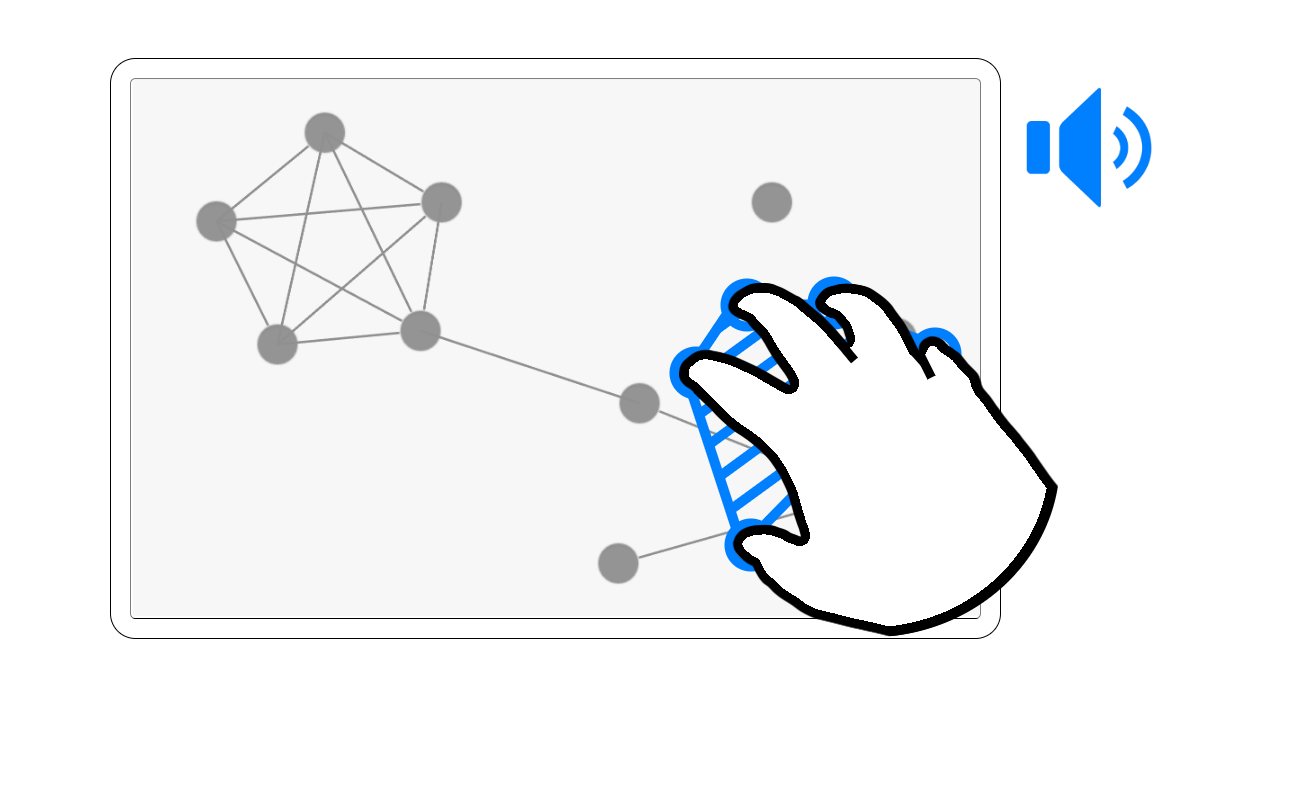}
        \subcaption{Five-finger Dome.}
        \label{fig:five_finger_dome}
    \end{subfigure}
    
    \begin{subfigure}[b]{0.39\textwidth}
        \centering
        \includegraphics[width=1\textwidth]{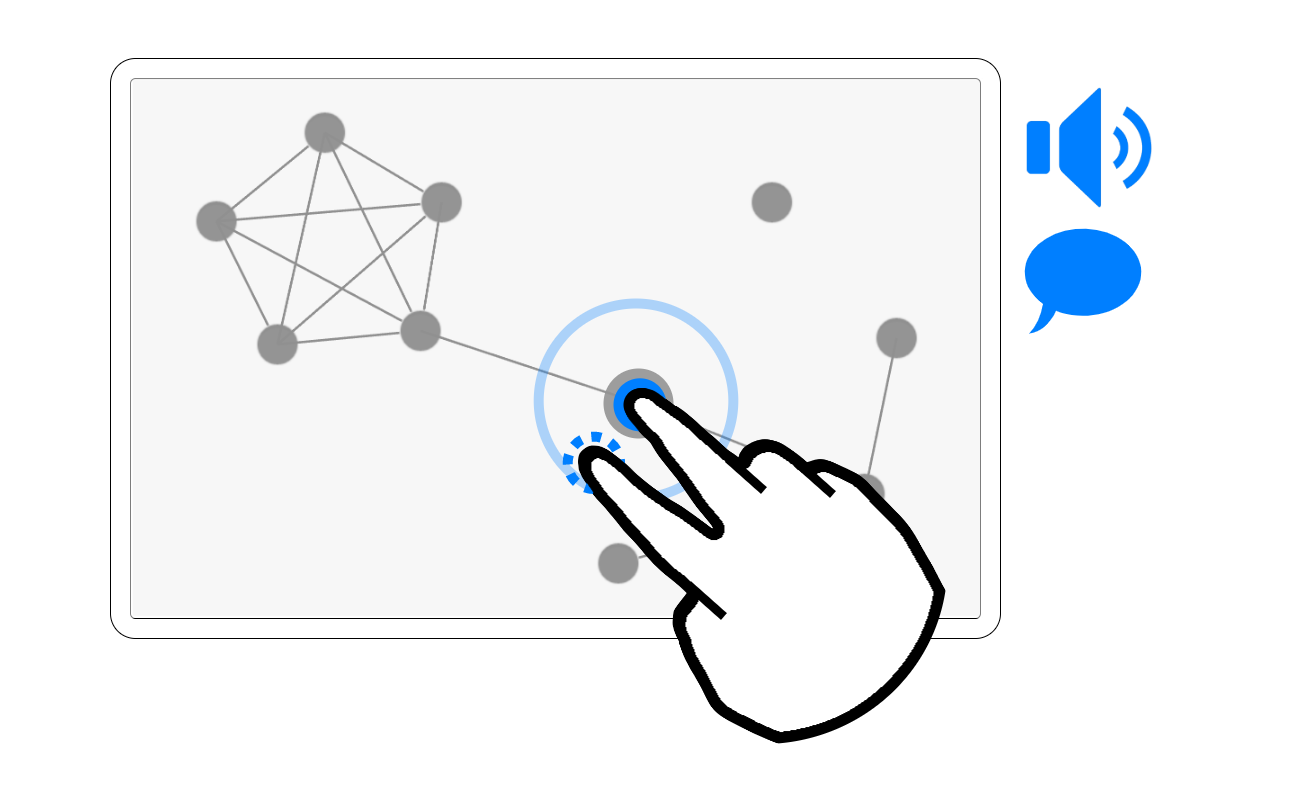} 
        \subcaption{Dwell+Tap with one hand.}
        \label{fig:dwell_tap_one_hand}
    \end{subfigure}
    \begin{subfigure}[b]{0.39\textwidth}
        \centering
        \includegraphics[width=1\textwidth]{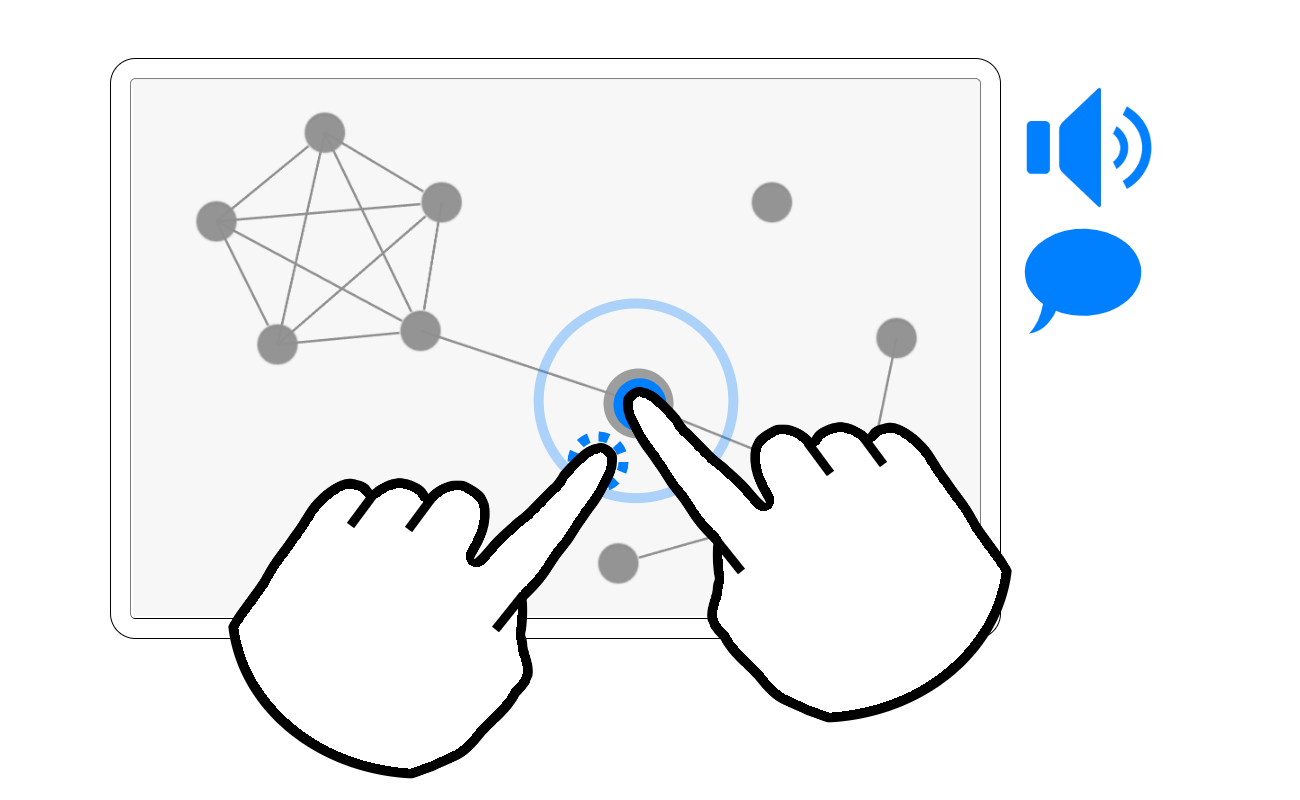}
        \subcaption{Dwell+Tap with two hands.}
        \label{fig:dwell_tap_two_hands}
    \end{subfigure}
    
    \begin{subfigure}[b]{0.39\textwidth}
        \centering
        \includegraphics[width=1\textwidth]{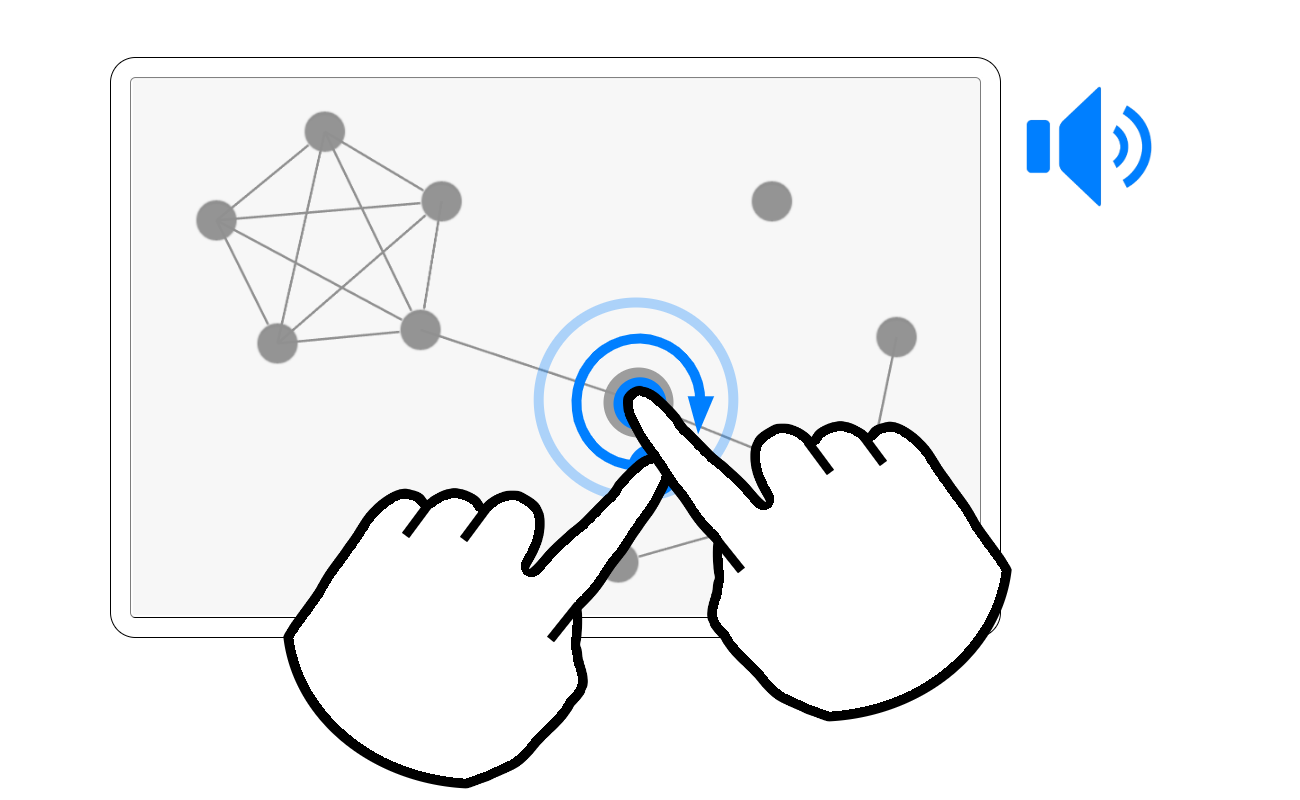} 
        \subcaption{Dwell+Circle.}
        \label{fig:dwell_circle}
    \end{subfigure}
    \begin{subfigure}[b]{0.39\textwidth}
        \centering
        \includegraphics[width=1\textwidth]{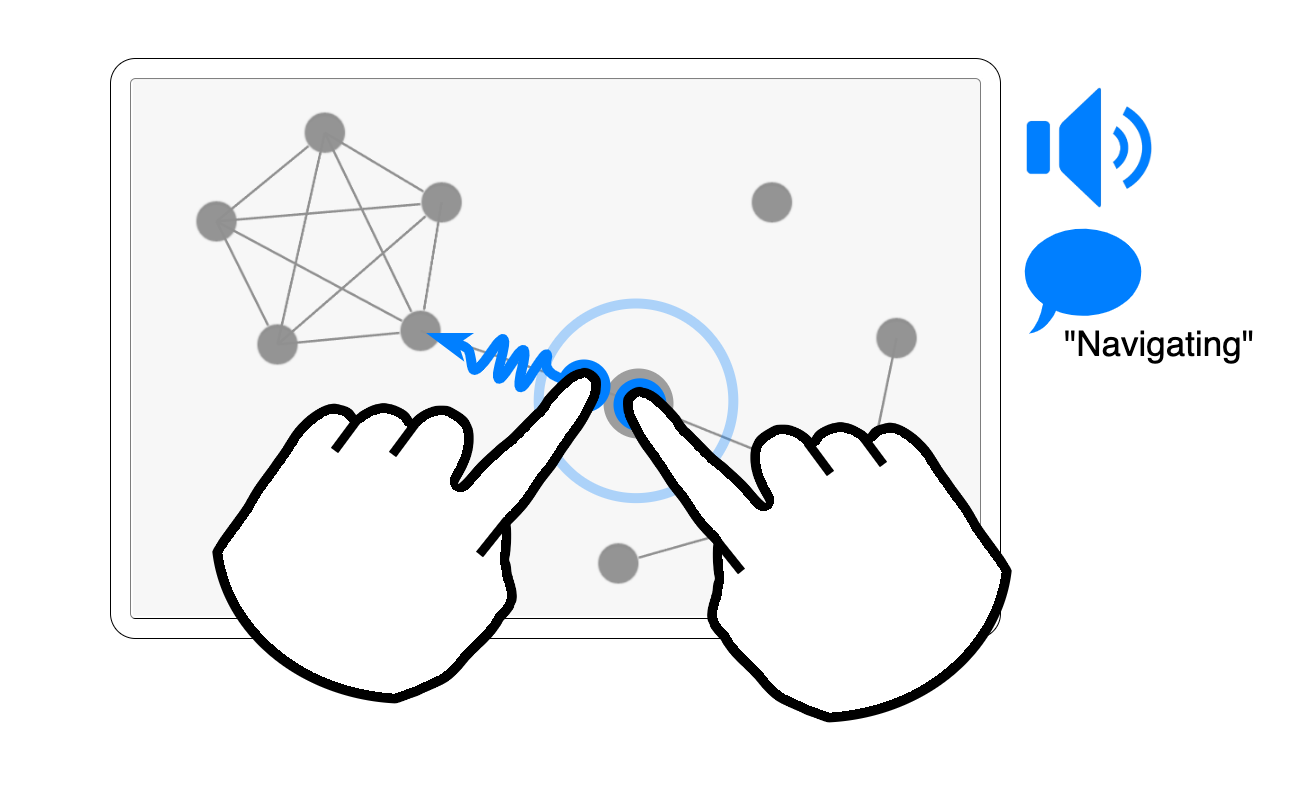}
        \subcaption{Dwell+Radiate.}
        \label{fig:dwell_radiate}
    \end{subfigure}

    \begin{subfigure}[b]{0.78\textwidth}
        \centering
        \includegraphics[width=0.9\textwidth]{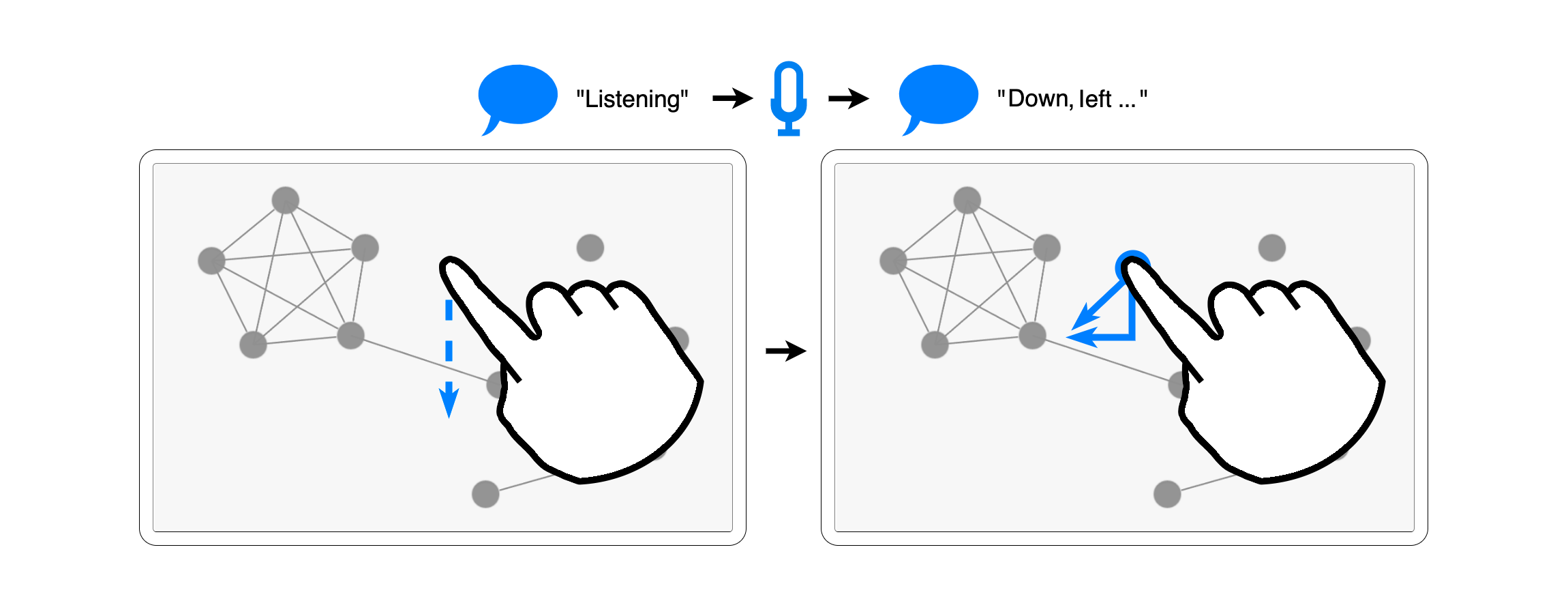} 
        \subcaption{Single-finger Flick-down, Speech Command for Searching and Single-finger Follow.}
        \label{fig:single_finger_flick_down_search}
    \end{subfigure}
    
    \caption{The seven interaction techniques supported by TADA. Continuous arrows indicate touch traces of the finger and discontinuous arrow are flicks. The speakerphone and speech bubbles indicate the use of abstract sounds and speech respectively.}
    \label{fig:interactions}
    \Description{The figure has seven sub-figures, each depicting an interaction technique TADA supports. Sub-figure a) displays a Single-finger Sweep with corresponding sounds. b) presents a Five-finger Dome forming an area on the screen, also producing sounds. c) demonstrates dwell and tap with one hand, where a finger stays on a node and another taps nearby, generating sounds and speech. d) depicts dwell and tap with two hands, using one finger from each hand. e) exhibits dwell and circle, where a finger stays on a node and another circles around it, producing sounds. f) portrays dwell and radiate, where a finger moves away to navigate towards a target node, producing audio and the spoken word "navigating". Finally, g) shows a Single-finger Flick-down initializing the microphone, and the system says "listening". The user instructs the system with speech input to search for the target in a diagram. TADA then produces speech instructions, such as "down, left", to direct the finger's movement. 
    } 
\end{figure*}

Our main way to instantiate \textbf{DP3} and \textbf{DP4} to achieve \textbf{G1} is to design a set of interaction techniques and pseudo-modes that support a variety of different ways to access the information contained in the diagram. One challenging aspect of the design process, which is difficult to articulate here in prose, was ensuring that the techniques are consistent and cohesive in both audio design and interaction. These techniques are intentionally crafted, whenever feasible, to work concurrently [\textbf{DP4}]. The Video Figure in the supplementary materials conveys some of these operations better than the overview (Figure~\ref{fig:interactions}). In the names of the interaction techniques below we use a ``+'' in the name when two fingers operate simultaneously within the same technique but with different functions. Whenever a technique was inspired by a previous system, or when there is previous empirical evidence about the interactions that work best, we incorporate that knowledge and mention the sources.

\paragraph{Single-finger Sweep:}
\label{single-finger-sweep}

As a finger sweeps over the screen, the nodes that are touched produce a discrete horn sound and the links that are crossed produce a discrete plucked string sound (Figure~\ref{fig:single_finger_sweep}). If the movement is faster, the sounds are shorter. If the finger stops on a node or a link, the sound is sustained indefinitely. A small degree of interaction hysteresis is applied in this case: if the movement is very slow, a node grows a bit to accommodate drift, assuming that the user just intends to keep a finger in that node.
In practice, the Single-finger Sweep enables quick overviews of the contents of the diagram or a particular area and some degree of detailed exploration. However, it is not appropriate for precise detail retrieval, since that would require systematic tracing of the whole surface, which proved difficult and error-prone in our experience.

Several existing systems have a similar interaction. For example, in AudioGraf~\cite{Kennel_1996} and PLUMB~\cite{Cohen_Yu_Meacham_Skaff_2005,Calder_Cohen_Lanzoni_Xu_2006,Cohen_Haven_Lanzoni_Meacham_Skaff_Wissell_2006,10.1145/1124706.1121428} a similar interaction technique is the only way to interact with the content.

\paragraph{Five-finger Dome:}
\label{five-finger-dome}

Users activate the Five-finger Dome by placing all the fingers of the hand on the tablet's surface, which defines a pentagonal area. This plays an audio stream of sounds configured in the following way: one node inside the dome area plays first (french horn), then its connections (string sounds), then another connected node (still within the dome area), with its connection, and so on until all nodes in the dome area are exhausted. Then a short bell sound acts as a separator, and the same sequence plays again (unless the dome has changed position, in which case the sequence changes after the next bell). This technique provides a more refined way to explore connection patterns in localized areas. It is designed to produce sound textures that are perceivably different depending on the connection patterns. For example, an area of nodes connected in a ring produces a homogeneous rhythmic alternation of node and link sounds, whereas a hub-and-spoke-like pattern produces a sequence that starts with a single node sound followed by many link sounds and then an alternation of node and link sounds. The duration of a full cycle is the same every time, so that denser areas produce faster sequences and vice versa; the overall duration of a cycle is configurable through the system settings.

This technique still supports overview, but in a more abstract way than the Single-finger Sweep, which would be difficult and tedious to use to recognize connectivity patterns. We also anticipated that the dome would require additional expertise to master. The technique is inspired by sonification systems, such as those that provide static renders of charts~\cite{Alty_Rigas_2005, Hansen_BaltaxeAdmony_Kurniawan_Forbes_2019}. However, to our knowledge, our system is the first to leverage higher-level sound textures interactively.

\paragraph{Dwell+Tap:}
\label{dwell+tap}

Once a finger dwells on a node or a link\footnote{Reaching a node can be achieved through Single-finger Sweep, described above, through Search and Follow, or by navigating the network, described below.}, a tap by a second finger in the proximity of the dwelling finger causes the system to read (through speech synthesis) the name of the object and, for each consecutive tap, its attributes. We moved away from Kane et al.'s introduction of ``Split Tap''~\cite{Kane_Bigham_Wobbrock_2008, Kane_Morris_Perkins_Wigdor_Ladner_Wobbrock_2011, lazar2015ensuring}, in which the second finger can tap anywhere, to enable multiple of these gestures simultaneously on several objects [\textbf{DP4}] and to make the system more robust against accidental taps~\cite{Mcgookin_Brewster_Jiang_2008}.

\paragraph{Dwell+Circle:}\label{dwell+circle}Once a finger is on a node, the user might want to know how many times and in which directions it is connected to other nodes. A second finger in proximity, orbiting around a finger dwelling on a node (which could have gotten there in a number of ways), triggers the Dwell+Circle technique. This stops the continuous sound from the dwell on the node and provides an additional tone that becomes louder as the finger approaches the angle of a link. As the link is crossed, it makes the regular string sound, and if the link is dwelt on, this can transition seamlessly to Dwell+Radiate below (in direct support of~\textbf{DP4} and consistent with Hinrichs and Carpendale's findings about touch gesture transitions in public displays~\cite{hinrichs_gesture_sequences}). The fingers do not have to be from different hands but, in certain situations two-touch rotation with a single hand becomes awkward (see~\cite{hogan_rotation_gestures}).

A specific technique to find links such as Dwell+Circle is not strictly necessary, since a second finger doing Single-finger Sweep could potentially enable finding the links around. However, Dwell+
Circle provides the following advantages: 1) it declutters the audio space by interrupting the node dwell sound; 2) it facilitates finding the links through the additional proximity guidance tone, or detecting right away that the node is not connected (absence of the tone); 3) it enables a seamless transition to the Dwell+Radiate technique; and, 4) it enables us to filter out nodes or links that are not relevant to the current node (e.g., close nodes, or links just crossing). We are not aware of a similar technique in any existing system.

\paragraph{Dwell+Radiate:}
\label{dwell+radiate}

After finding a node and a link of interest, the user will often want to follow the link to the corresponding neighbour node. This is a common form of navigation through the network. While one of the fingers still dwells on the node and the other finger is on a link (from the preceding Dwell+Circle gesture), the user can start moving this second finger away from the origin node (instead of the circular motion in Dwell+Circle), which triggers Dwell+Radiate. The tone sound now becomes louder as the finger approaches the neighboring node and plays a fanfare twist of the french-horn sound of the node to indicate success, at which point the original dwelt node finger can be released to start a new Dwell+Circle gesture. There are two main supports for the user tracing the line to the new node. First, the proprioceptive information of the two fingers can help the user move one finger away from the other (similar to a pinch gesture). Second, the string sound stops playing if the finger deviates from the string, at which point the user can seek it again by moving perpendicularly up and down. The design of this technique is inspired by existing work on tracing lines~\cite{Tennison_Gorlewicz_2019, Gorlewicz_Tennison_Uesbeck_Richard_Palani_Stefik_Smith_Giudice_2020}, although it introduces new elements (e.g., filtering out other links, seamless transition to further navigation) and it is seen here for the first time together with other forms of interaction. 

\subsection{Pseudo-modes}

Besides the interaction techniques described above, TADA offers several pseudo-modes, often triggered through speech commands, that support different kinds of access. A Single-finger Flick-down invokes the tablet's speech recognition system, which responds with ``listening''. Recognized commands are Searching and Filtering, which we discuss below.

\paragraph{Searching with Single-finger Follow:}
\label{searching}

After a Single-finger Flick-down, saying ``Search for [information]" invokes the search mode. The user can search for information contained in any node, link, or for the links or node names themselves. For example, in a social network diagram, the user can flick down and utter ``searching for Asha''. The system repeats the speech command from the user and processes the search. The system responds verbally with the results, for example, it could say ``Nothing found", or ``Found [number of results] result(s)" depending on the search. To locate the results, the user places a single finger on the screen. The system then provides directions (left, right, bottom, up) verbally that direct the user's finger to the location of the nearest result. For example, if the node for ``Asha'' is in the lower left direction relative to the finger, the system would say ``down, left" on a loop. The verbal directions adapt to the movement of the finger. If the finger moves diagonally, the directional instruction combines horizontal and vertical prompts. Alternatively, when the finger moves only horizontally or vertically, the system will provide only directions in that axis until the horizontal or vertical location of the target is reached, and then the other directions are provided. The pacing of the speech prompts also encodes the remaining distance to the target. A slower pacing means the finger is closer. When the finger reaches a target node the system plays the same fanfare-like horn sound at the end of Dwell+Radiate. 

We considered several ways to guide users to a target. We incorporate Ramôa et al.'s approach, which found that voice-based guidance is the most efficient method for helping people pinpoint a location, and it does not require prior training~\cite{Schmidt_2022}. Similar techniques are utilized in ImageAssist~\cite{Nair_Zhu_Smith_2023} and AccessOverlays~\cite{Kane_Morris_Perkins_Wigdor_Ladner_Wobbrock_2011}, but without the axial and proximity enhancements.

\paragraph{Filtering:}
\label{filtering}

The same Single-finger Flick-down triggers a filtering operation if the user says: ``Filter by [attribute, value]". The system recognizes the end of the speech command, repeats the command back and enters into filtering mode. 
The filtering mode reduces the complexity of a diagram by deemphasizing nodes and links that do not comply with the filter condition (a value of an attribute). For example, one can filter in a genealogical diagram by attribute gender: female. The attributes and values that users can filter for are the same ones accessible through the Dwell+Tap technique. When the filter is on, only the nodes that meet the filter criterion and their connections are fully accessible, but to preserve some degree of awareness of the filtered-out elements, these nodes and their connections still produce sounds, but at a fainter volume and with a static white noise played on top of them. This signifies that these elements have been filtered out and are not currently the focus of the person’s attention.

\paragraph{Additional Information:}
People can listen to the alt-text of the diagram by flicking a single finger to the right. Flicking with two fingers to the left plays the audio legend: a description of the audio mappings (e.g., plays the horn sound and says ``node'').

\bigbreak 

Overall the techniques and pseudomodes intend to bring people from \textbf{L3} up (we do not address \textbf{L1}-\textbf{2}). For example, Single-finger Sweep, Five-finger Dome and Dwell+Circle all offer multiple [\textbf{DP3}], complementary [\textbf{DP2}, \textbf{DP4}] and spatially-indexed access [\textbf{DP1}] to locations decided/remembered by the user, enabling people to reach \textbf{L4}. Moreover, Dwell+Tap, Dwell+Circle and Dwell+Radiate are designed to alternate each other efficiently, bringing people closer to \textbf{L5}.

\section{Evaluative Study}
\label{sec:evaluation}

The main goal was to test whether participants could use the fundamental features of TADA effectively and to detect usability issues. This formal study complemented the continuous informal testing of the system during design and implementation that we carried out on ourselves and others, including a blind co-author. 

\subsection{Participants}

Twenty-five participants (14 female, 10 male, 1 transgender) who self-identified as legally blind and were 25 to 101 in age (mean=51.92, median=52, SD=16.35) volunteered for the study. Because visual perception can vary significantly, we analyze by group: group 1 (20 participants) could not visually access diagrams on the tablet at all; group 2 (5 participants) could perceive visual information from diagrams to a varying degree. The specifics of participants' demographic information are in Appendix~\ref{appendix:evalStudy:ParticipantDetails}. Participants received CAD\$30 compensation. This study was pre-approved by the local REB.

\subsection{Location and Apparatus} 
Experiments took place in 3 different locations, all dedicated private spaces with a table. 
TADA ran on a 10.6" Lenovo Tab M10 Plus tablet with Android 12 that we augmented with a thin 3D-printed bezel to prevent participants from inadvertently sliding their fingers off its edge. We used a 3D-printed tactile overlay of a simple diagram to speed up training. We captured the hands of the participant from above with a tripod-mounted video camera.

    


\subsection{Procedure and Measurements}
Participants provided consent and verbally answered demographic questions. After an introduction to the system's purpose, participants completed six blocks of activities. Each block consisted of a training phase with the tactile overlay and without it and then interaction with the tablet to answer questions or complete task requests. When the participant asked an additional question or required further assistance, the experimenter recorded the task as ``completed with assistance''. We designed the experiment's diagrams to be representative of reasonably complex diagrams that one might encounter in real life but also to have unequivocally correct or incorrect answers (see also scalability discussion---Subsection \ref{sec:discussion:TADADesign}). The diagrams used for training and to answer the questions were different. After each block, the experimenter requested general comments about the interaction techniques. After all tasks, participants verbally answered a NASA TLX questionnaire. We changed the original TLX questionnaire scale from 21 to 10 points to make it more convenient for verbal interaction. 


We collected: a) demographic data; b) correctness of answers or success in activity; c) comments on interaction techniques; d) video of the tablet and hands; e) experimenter's observations and; f) NASA TLX responses. The experiment took 45-90 minutes.

\subsection{Tasks and Stimuli}
We designed six experimental task blocks, each testing an interaction technique or basic functionality of the system. For example, the first block tests participants' ability to achieve an overview with the Single-finger Sweep technique of Section~\ref{single-finger-sweep}. 

\begin{figure}[htbp]
    \centering
    \begin{subfigure}[b]{0.23\textwidth}
        \centering
        \includegraphics[width=1\textwidth]{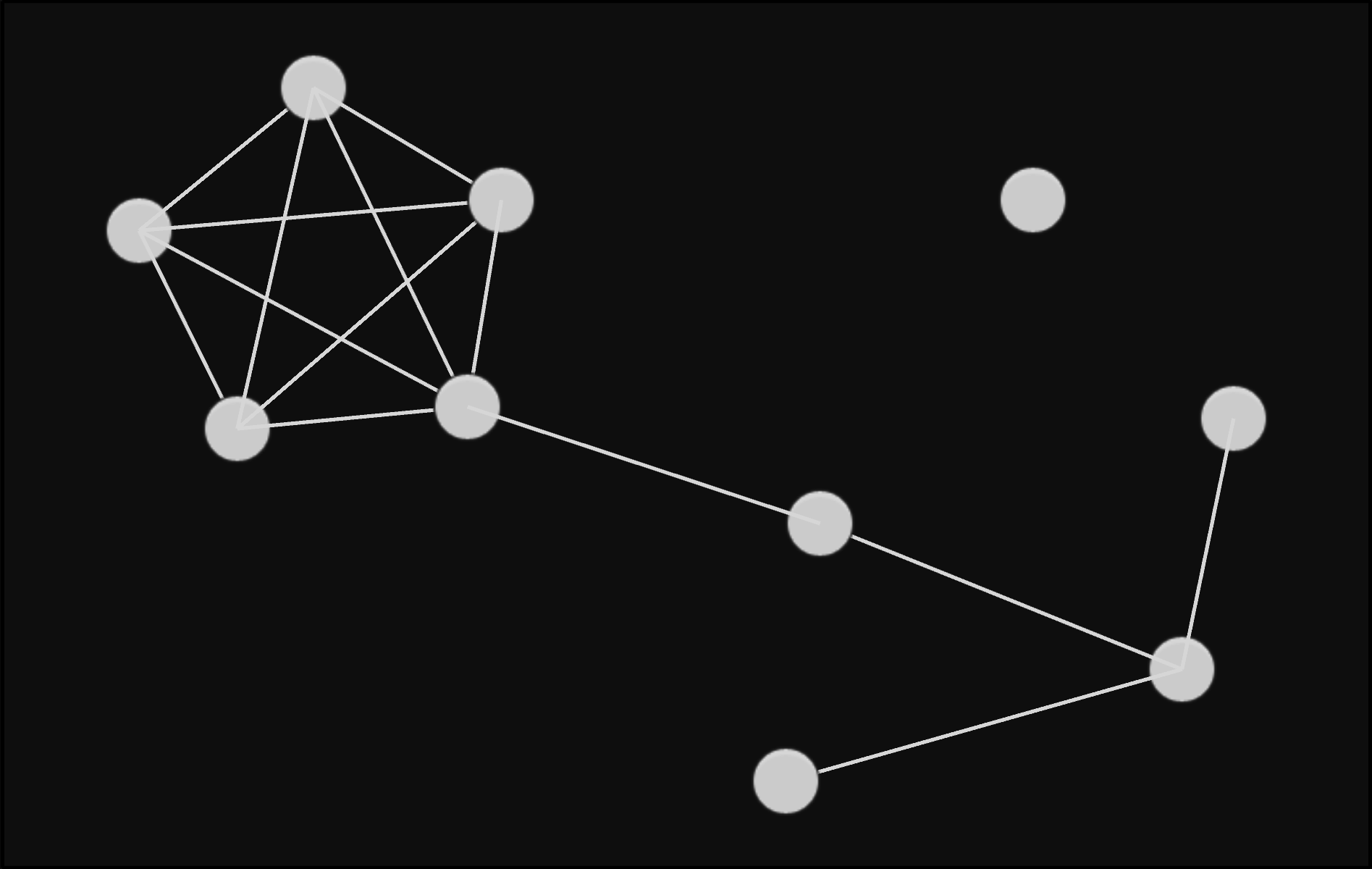} 
        \subcaption{Diagram for learning.}
        \label{fig:device_diag_leaning}
    \end{subfigure}
    \hspace{0.003\textwidth}
    \begin{subfigure}[b]{0.23\textwidth}
        \centering
        \includegraphics[width=1\textwidth]{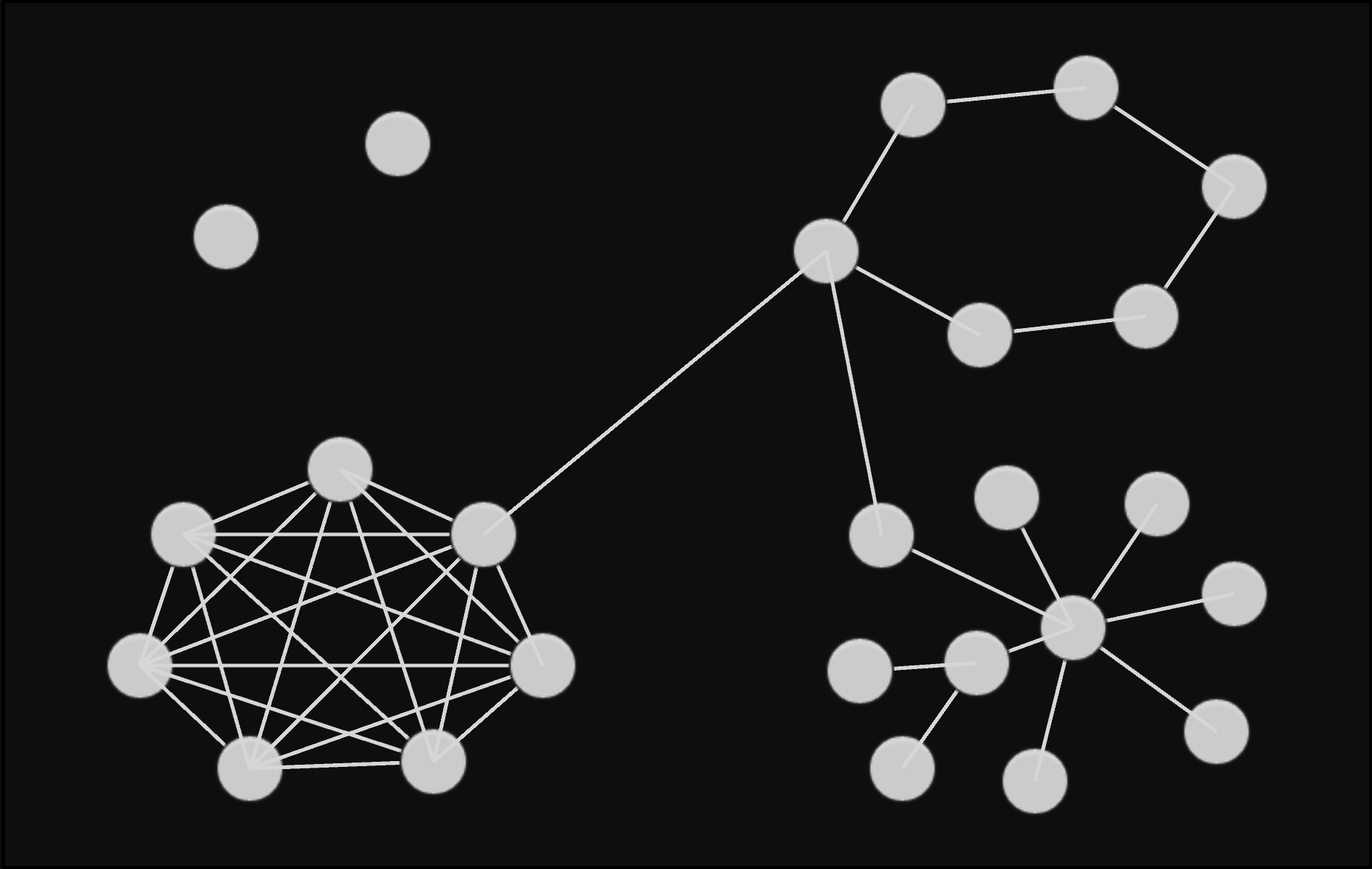}
        \subcaption{Stumuli diagram 1.}
        \label{fig:device_diag_1}
    \end{subfigure}
    \begin{subfigure}[b]{0.23\textwidth}
        \centering
        \vspace{0.05\textwidth}
        \includegraphics[width=1\textwidth]{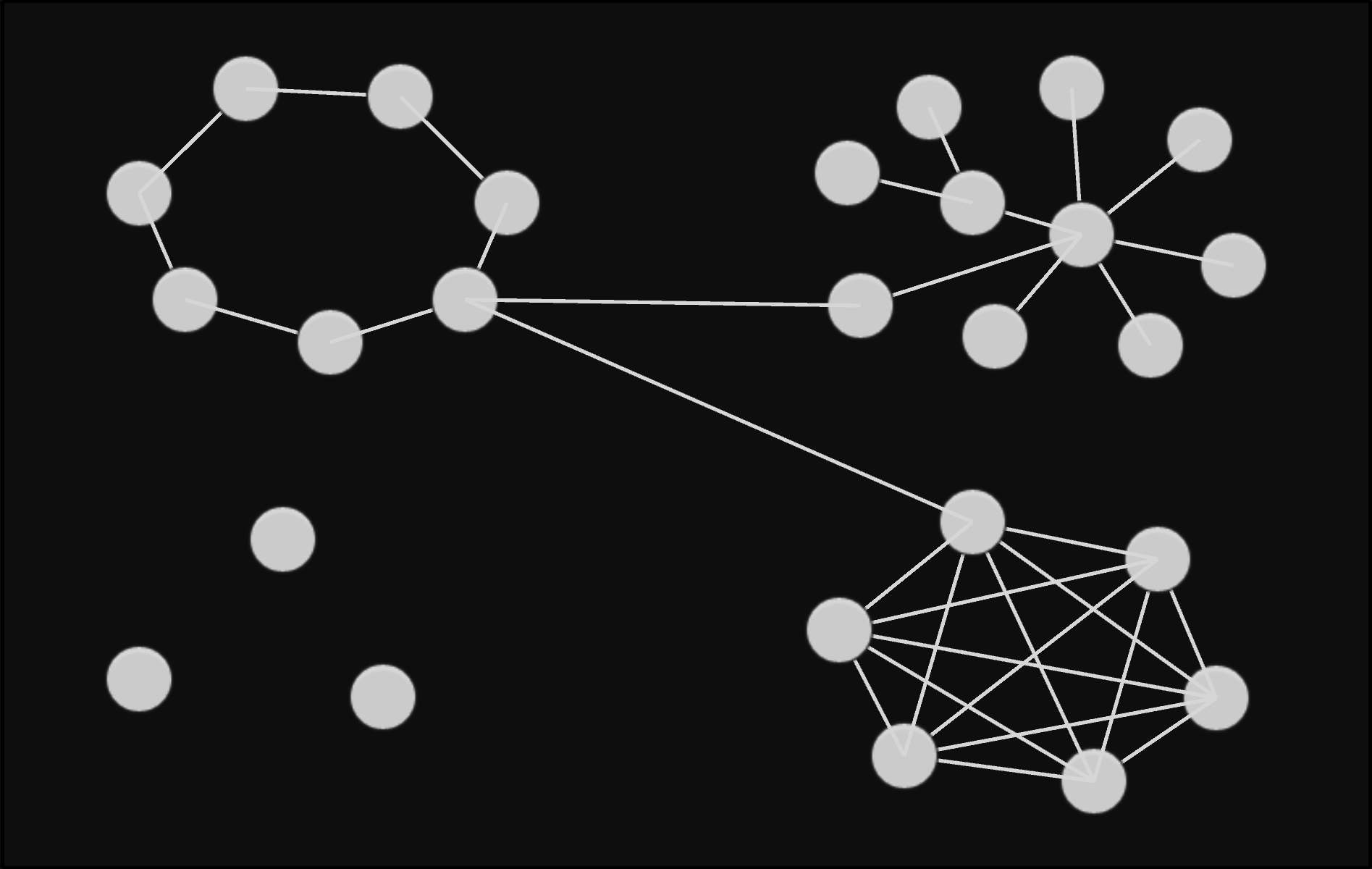} 
        \subcaption{Stumuli diagram 2.}
        \label{fig:device_diag_2}
    \end{subfigure}
    \hspace{0.003\textwidth}
    \begin{subfigure}[b]{0.23\textwidth}
        \centering
        \vspace{0.05\textwidth}
        \includegraphics[width=1\textwidth]{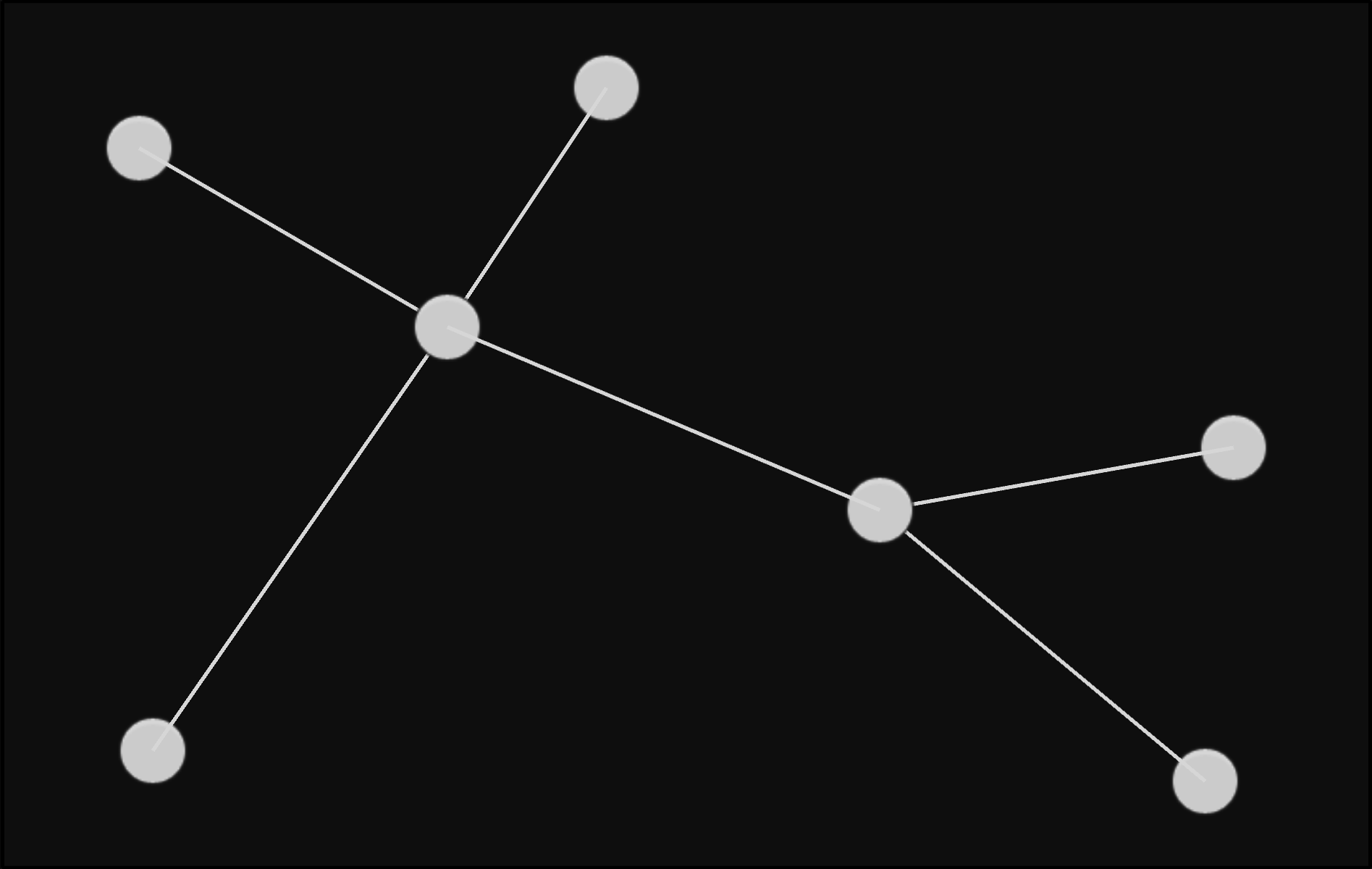}
        \subcaption{Stumuli diagram 3.}
        \label{fig:device_diag_3}
    \end{subfigure}
    \caption{Diagrams presented in TADA as stimuli.}
    \label{fig:device_diagrams}
    \Description{There are four sub-figures, each portraying a different diagram used in the evaluative study. Sub-figure a) shows the learning diagram, in which there are people who are friends with each other showing high connectivity and forming a cluster on the left, with an empty space under this cluster. One person from the cluster has a friendship extending to the lower right region which is more sparse, with fewer people and friendships. There is one person on the upper right not connected to anyone.  
    b) shows test diagram 1. The first quadrant (upper left) has few people with no friends. The second quadrant (upper right) has more people, and each is a friend with two other people in the same quadrant. The third quadrant (lower left) has even more people who are friends with each other, having the highest number of connections. The fourth quadrant (lower right) has the most number of people, with one popular person having high connections, and others who have only one or two friends. Quadrant 2 also connects to quadrants 3 and 4. 
    Test diagram 2 from c) included modifications to test diagram 1 such as changing the positions and numbers of the nodes in each quadrant. Specifically, we added one more node to each of quadrants 1 and 2, removed one node from quadrant 3 and flipped the node positions in quadrant 4 in test diagram 1 while keeping the general spatial structure for the evaluation. We then shuffled the quadrants so that quadrants 1, 2, 3 and 4 from test diagram 1 now correspond to quadrants 3, 1, 4 and 2 in test diagram 2. 
 Test diagram 3 from d) only has 7 nodes sparsely connected. Towards the center slightly to the left, there exists the most connected node.}
\end{figure}

The design of the tasks (Table~\ref{tab:tasks}, Appendix~\ref{appendix:evalStudy:TaskDetails}) was informed by our formative study (tasks mentioned by participants) and prior work~\cite{Yang_Marriott_Butler_Goncu_Holloway_2020, Lee_Plaisant_Parr_Fekete_Henry_2006}. These cover many tasks prevalent in node-link diagram research such as overview, clusters, adjacency, and connectivity aspects. Overview refers to the full diagram, whereas summary can be just of certain areas. We selected the experimental tasks to balance: a) being representative of a wide variety of common tasks; b) having quantifiable outcomes; and c) assessing one or more of the interaction techniques offered by TADA. 
%
We created the stimulus diagrams (Figure~\ref{fig:device_diagrams}) to meet the goals and constraints of our experiment. We prioritized simple layouts to make the experiment manageable and easy to explain, to avoid confusion over required outcomes and to reduce time to achieve familiarity. Stimuli 1 and 2 contained multiple networks with different configurations within the same diagram so that we could evaluate their ability to contrast and compare quickly without excessive memory strain. 

\begin{table*}[htp]
  \centering
  \caption{Summary of the six experimental blocks.}
  \Description{This table lists the questions or tasks required for participants to answer or perform for the interaction technique evaluated in each experimental block. }
  \resizebox{\textwidth}{!}{%
    \begin{tabular}{llcl} 
    \thead{Block} & \thead{Technique} & \thead{\#Q}   & \thead{Example Question/Action} \\
    \midrule
    Overview & Single-finger Sweep & 5     & \multicolumn{1}{p{35.00em}}{Q1.1: Describe the overall layout of the diagram.\newline{}  
                                          Q1.2: Which quadrant of the diagram has the most connections between people?\newline{}
                                          Q1.3: Which quadrant has the least connections?\newline{}
                                          Q1.4: Which quadrant of the diagram has the most people?\newline{}
                                          Q1.5: Which quadrant has the least people?}\\\midrule
    Summary & Five-finger Dome & 5     & \multicolumn{1}{p{35.00em}}{
                                          Q2.1: Which quadrant of the diagram has the most connections between people?\newline{}
                                          Q2.2: Which quadrant has the least connections?\newline{}
                                          Q2.3: Which quadrant of the diagram has the most people?\newline{}
                                          Q2.4: Which quadrant has the least people?\newline{}
                                          Q2.5: Which quadrant has people with unequal numbers of friends?}\\\midrule
    Detail & Dwell+Tap & 1     & Q3: Given a person, what is the name and professional background associated with it? \\\midrule
    Search & \multicolumn{1}{p{21em}}{Flick Down, Speech Command, Single-finger Follow} & 1     & Q4: Locate Bob. \\\midrule 
    Connections & Dwell+Circle & 1     & Q5: How many friends does Bob have? \\\midrule
    Navigate & Dwell+Radiate & 1     & Q6: Navigate from Bob to any of his friends.  \\\midrule
    \end{tabular}%
    }
  \label{tab:tasks}%
\end{table*}%


\subsection{Data Analysis}

We employed a mixed-methods approach to analyze the collected data. All the interview responses were analyzed qualitatively using a thematic analysis approach~\cite{Adams_Lunt_Cairns_2008, boyatzis1998transforming}, wherein we first used deductive coding and then organized the codes into themes using affinity diagramming. For questions Q1.2-Q6 we also counted the number of correct and incorrect responses. We referred to the gestures video to better understand participants' feedback regarding challenges and common causes of error but did not perform a systematic gesture analysis which we leave to future work.



\subsection{Results}

This section summarizes the quantitative and qualitative results of our study.

\subsubsection{Task and Question Outcomes} 

\begin{figure}[htp]
    \centering
    
    \begin{subfigure}[b]{0.49\textwidth}
        \centering
        \includegraphics[width=1\textwidth]{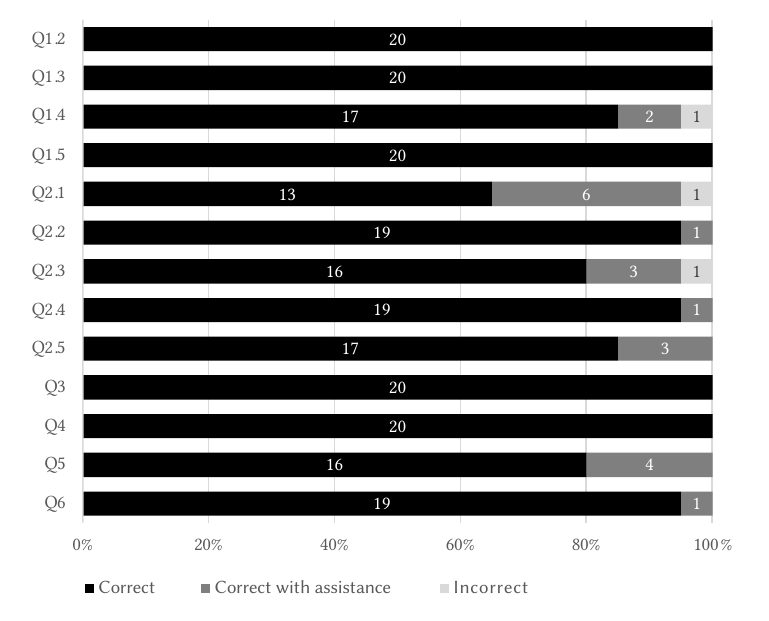} 
        \subcaption{Group 1 (no visual diagram perception) results.}
        \label{fig:group1stats}
    \end{subfigure}
    \begin{subfigure}[b]{0.49\textwidth}
        \centering
        \includegraphics[width=1\textwidth]{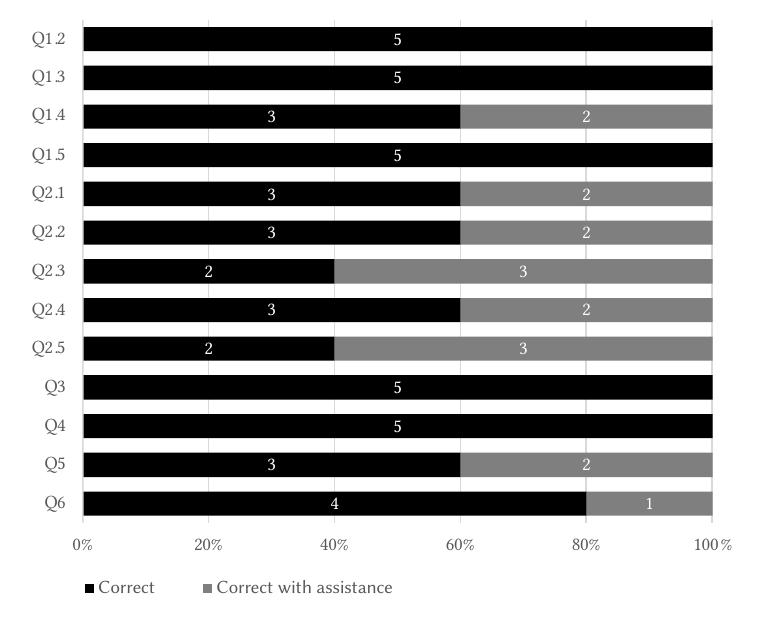}
        \subcaption{Group 2 (partial visual diagram perception) results.}
        \label{fig:group2stats}
    \end{subfigure}
    
    \caption{Correctness of answers to quantitative questions listed in Table \ref{tab:tasks}.} 
    \label{fig:task_stats}
    \Description{The figure presents participant answer accuracy as the x axis for groups 1 and 2, categorized as correct, correct with assistance, or incorrect from left to right. The y axis is the list of questions. In general, we see that the figure is mostly occupied by correct responses from participants. Sub-figure a) for group 1 shows most responses were correct. Q1.4, 2.1, 2.3, 2.5, 5 and 6 needed assistance (1 to 6 instances in a question), and only 3 instances were incorrect. Sub-figure b) for group 2 shows that a majority were correct (though less percentage than group 1). The rest of the responses were assisted but they were also all correct. }
\end{figure}

Figures~\ref{fig:task_stats}a and b present a summary of participants' responses to Q1.2-Q6. A majority of the participants from both groups answered all the questions correctly (22 out of 25) and 7 out of 25 participants required assistance with more than one question, such as additional clarifications regarding the underlying technical concepts. Lastly, there were three instances of three participants who did not reach the correct answer. 

For Q1.1, all participants from both groups could correctly describe the overall layout of the diagram. Three participants [P1, P8, P14] described the layout by explaining how the nodes were spatially organized. For example, P8 from group 2 said: \textit{``That's a group of people [on the lower left ... There are] single individuals [on the upper left].''} Our expectation when designing TADA was that participants would also be able to identify details such as connectivity between regions and nodes, and we found this to be the case; 22 out of 25 participants from both groups described the diagram at a richer level. For example, P6 from group 1 described it as \textit{``[On the lower right,] there is a person in the middle, I think, that's connected to all other people around ... as I am going in circle here.''}

To address Q1.4-Q6, we occasionally provided additional clarifications or assistance to help the participant better understand the questions or to complete the task. For instance, Q1.4 and Q2.3 require identifying the quadrant with the most nodes. Three participants in Q1.4 and 2 participants in Q2.3 believed that the quadrant having the most connections is the one with the most nodes. To clarify the question, we described the difference between the density of nodes and the density of links, and this helped the participants correctly identify the answers. Another instance of such clarification involved prompting participants to think more carefully about the question. For example, to answer Q2.5 (people with unequal numbers of friends), three participants provided an answer but we recognized that were unsure about the underlying technical concept, and therefore, we prompted them to think more carefully about what the question was asking and then they answered correctly. Six participants required assistance with their gesture(s), such as covering a quadrant more completely with the Five-finger Dome interaction. 

Three participants from group 1 did not correctly answer Q2.1 (most connections in a quadrant),  Q1.4 and Q2.3 (most nodes in a quadrant). They confused the region where only one node was highly connected with the region where all nodes were connected to each other. We do not think this was due to misinterpretation of the audio because they acquired correct answers for all remaining questions.

\subsubsection{Perceived Workloads}

\begin{figure}[htp]
    \centering
    
    \begin{subfigure}[b]{0.47\textwidth}
        \centering
        \vspace*{0.1cm}
        \includegraphics[width=1\textwidth,trim={0 3.8cm 0 0.5cm},clip]{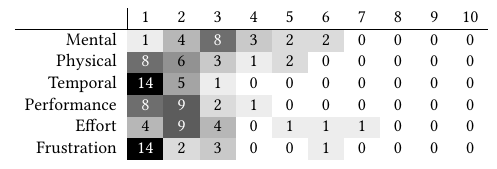} 
        \subcaption{Group 1 NASA-TLX results.}
        \label{fig:group1tlx}
    \end{subfigure}
    \begin{subfigure}[b]{0.47\textwidth}
        \centering
        \vspace*{0.2cm}
        \includegraphics[width=1\textwidth,trim={0 0.1cm 0 4.2cm},clip]{./figures/evalStudyStats/NASATLX.pdf}
        \subcaption{Group 2 NASA-TLX results.}
        \label{fig:group2tlx}
    \end{subfigure}
    
    \caption{NASA-TLX workload rating histograms. The number in each cell indicates the number of participants selecting that workload rating.}
    \label{fig:TLX_hist}
    \Description{These are the NASA-TLX questionnaire ratings for groups 1 (left) and 2 (right). Cells in rows labelled “Mental”, “Physical”, “Temporal”, “Performance”, “Effort” and “Frustration” and columns numbered 1-10, shows how many participants rated that aspect at that numerical level. In both tables, lower ratings have higher frequencies. A higher frequency is coloured more intensely on a greyscale. For both groups, the workload ratings are generally towards the left side of the tables, meaning the workloads were generally considered low. }
\end{figure}

Figure~\ref{fig:TLX_hist} summarizes participant responses to the NASA TLX questionnaire~\cite{Hart_Staveland_1988, Hart_2006}. 
Overall, we visually infer from Figure \ref{fig:TLX_hist} that the workload ratings are generally towards the left side of the tables (i.e., the workloads were generally considered low). Mental demand (group 1's mean rating: 3.35; group 2: 3.8), physical demand (1: 2.15; 2: 4.6) and effort (1: 2.7; 2: 3.2), received slightly higher mean ratings from both groups than other categories, for which the mean ratings were below 2. Participants who gave higher scores in these categories reasoned that the cognitive effort needed to understand and interpret the audio information was relatively high (n=4) and so was the physical effort for performing some of the gestures (n=2). P8 explicitly attributed the workload to the effort required to answer the questions due to their age (101 years). In the cases of physical demand, this was due to a lack of experience with performing gestures [P8] and low dexterity of fingers [P12].

\subsubsection{Participant Feedback}

Overall, participants described TADA as \textit{``fun"} [P2, P11, P16], \textit{``intuitive"} [P3, P4, P5, P22], \textit{``useful"} or \textit{``helpful"} [P1, P3, P4, P5, P7, P11] and \textit{``innovative"} [P1, P18, P22, P25].

\paragraph{Audio Design}
Eight participants explicitly noted that associating distinct sounds to nodes and links was a good strategy. P5 complimented: \textit{``I like the distinction of the sounds. You know, you got a very distinct pluck with the string, and then the horn honks. [And] the navigational sound [of] orientation [is] a good contrast with the other two.''}. P10 found the use of speech and non-speech sounds for different purposes effective: \emph{``When you are just gathering a sense of the general layout, I think the sounds are perfect because if [the system] starts to tell you [such information in speech] all at once, you cannot distinguish [the information]"}. 

Participants found that the audio they received when performing specific interaction techniques was helpful. Participants found the audio feedback when performing the Single-finger Sweep was informative to gain an overview of a diagram (n=16). The paced link sounds heard when using the Five-finger Dome gesture helped many understand the ways in which nodes were connected within a sub-region of the diagram (n=11). Four participants explicitly noted that speech was appropriate to convey detailed information using the Dwell+tap interaction. Lastly, 3 participants explicitly noted that the directional feedback when performing the Single-finger Follow gesture was logical, helpful and efficient. 

Participants also pointed to some limitations of our audio design. Three participants [P2, P10, P11] mentioned that the audio summary from the Five-finger Dome interaction (Block 2) was overwhelming and difficult to process. In Q2.1, five participants confused the connected links sound of one node as referring to information about the whole quadrant. Four of these participants gave the correct answer after we reminded them of how the audio was designed.

\paragraph{Interaction Techniques}
\label{sec:eval:results:techniques}



%
Participants provided mixed feedback on the interaction techniques. 
For instance, P9 and P19 mentioned that the Five-finger Dome (Block 2) gesture was easy to perform. For the Dwell+Circle (Block 5) gesture P14, P23, and P24 found it easy to perform and P3, P5, and P18 said that it provided clear indications of the directions of connected links. Seven participants noted Dwell+Tap (Block 3) was simple, and P6 especially liked the “stability” and “consistency” of repeated tapping during dwelling. As for Dwell+Radiate (Block 6), P3, P16, and P23 found it to be straightforward and easy to use. 


Some participants needed assistance with performing the gestures, especially the Five-finger Dome and the Dwell+Circle. Three participants [P8, P13, P16] commented that the Five-finger Dome gesture felt awkward to perform. We think this was partly due to this gesture requiring participants to maintain consistent contact by applying a certain level of pressure. Six participants also found Dwell+Circle awkward to perform. One reason for this was that the dwelling finger was observed to be sometimes drifting as participants performed the circling gesture, resulting in accidental interactions with other nodes or links. Additionally, because the node sound is muted during this interaction to prioritize the proximity guidance audio, participants also did not realize that their finger had drifted unless they lifted it up and placed it down to confirm that the dwelling finger was still on the node. 

Five participants [P2, P4, P11, P12, P25] attributed their difficulty with performing gestures to individual experiences. For example, they highlighted that varied motor skills [P2, P11, P12] can challenge touch-based interactions. P2, P4, and P24 noted that more familiarity with the device’s dimensions and gesture detection could enhance interaction. Lastly, three participants also had long nails that made it difficult to perform the gestures precisely or comfortably.
\section{Discussion}\label{sec:discussion}

This Section addresses the evaluation study, the system design and its applicability.

\subsection{Discussion and Limitations: Evaluative Study}
Most participants successfully completed the experimental task blocks, with some occasional assistance. Our results highlight that a majority of the participants could explore the spatial layout (\textbf{DP1}---participants correctly answered spatial questions) and distinguish between the different graph elements (\textbf{DP2}---participants correctly associated sounds and speech to different elements to answer questions about nodes, links and information contained therein). Additionally, each interaction technique and pseudo-mode enabled participants to gain different perspectives about the diagram (\textbf{DP3}---participants correctly identified different levels of information from overview to detail, from global to local and from spatial to semantic). 


Designing accessible gestures is a complex problem. Our design aligns with existing guidelines towards this goal such as Kane et al.'s~\cite{Kane_Wobbrock_Ladner_2011}, who recommend multi-touch gestures for BLV people and suggest allowing approximate targeting methods~\cite{Kane_Wobbrock_Ladner_2011}. Providing audio cues for targets can also help~\cite{Gorlewicz_Tennison_Uesbeck_Richard_Palani_Stefik_Smith_Giudice_2020}. We learned about two additional low-level issues: feedback about the required amount of pressure people have to apply is often required (e.g., this was a challenge for the Five-Finger Dome), and dwell-based gestures suffer from finger drifting (e.g. with Dwell+Circle). Further exploring how such challenges can be addressed will help in designing accessible gestures.

However, there are also areas that require further investigation. While TADA's design addresses \textbf{DP4} by enabling simultaneous gestures, our tasks did not evaluate this scenario. We also did not perform systematic analyses of the video recordings to investigate details of gesture performance. In future, we will carry out a detailed gesture analysis of the video and a study that tests the benefits of more complex interaction chains.
Moreover, our stimuli were relatively simple and constrained by the study's goals. An ``in-the-wild'' exploration of diagram access can advance our understanding of situational aspects and the scalability of our approach.


\subsection{Discussion and Limitations: TADA's Design} 
\label{sec:discussion:TADADesign}

Our formative study gathered evidence supporting the importance of making diagrams accessible to BLV people and exposed the challenges and frustrations of accessing this kind of information.  We also learned that it is important to make information available at least at \textbf{L3}---static descriptions. This finding aligns with current academic and commercial solutions that focus on improving accessibility to visual elements by encouraging alt-text and improving image descriptions~\cite{Edwards_Gilbert_Blank_Branham_2023, Guidelines-Image-Descriptions}. However,  textual descriptions are often not enough because people require different information depending on their circumstances. Our formative interviews encouraged us to facilitate access to multiple perspectives of a diagram [\textbf{G1}]. This can be achieved in different ways. For example, one could ask a human (a common workaround) or a large language model to retrieve a particular aspect of a diagram (e.g., the number of nodes, whether it is densely connected). Some available systems such as Bing already enable this kind of interaction\footnote{E.g., Bing: \url{http://www.bing.com}}. Nevertheless, diagrams are often created because they offer a better representation of the subject matter than text; for example, for representing structures that are inherently not linear. Moreover, we often access diagrams not to obtain a piece of information, but to remember topics or to build our understanding and mental models~\cite{Larkin_Simon_1987}. TADA was designed under the assumption that spatial interactive engagement with a diagram is as beneficial for BLV people as it is for sighted people. Thus, we encoded this as \textbf{DP1}. The comments from the formative interviews and the results from the evaluation study show early support for our proposed approach.

The current implementation of TADA chiefly uses touch input (without tactile feedback) and audio output to provide interactivity. However, other modality choices are possible, such as vibrotactile output (e.g.,~\cite{Poppinga_2011, Simonnet_2012}) and pin arrays~\cite{Vidal-Verdu_Hafez_2007}, which are progressively becoming more affordable and higher resolution. We think that some of these modalities offer advantages for some tasks. For example, following a link as done in our study might be much faster with a tactile interface, especially if challenges around resolution of the tactile display (e.g., pin arrays) can be improved. Exploring such solutions and comparing the benefits and limitations of different modalities may be useful for the design of future interfaces.

\begin{figure}[ht]
    \centering
    
    \begin{subfigure}[b]{0.23\textwidth}
        \centering
        \includegraphics[width=1\textwidth]{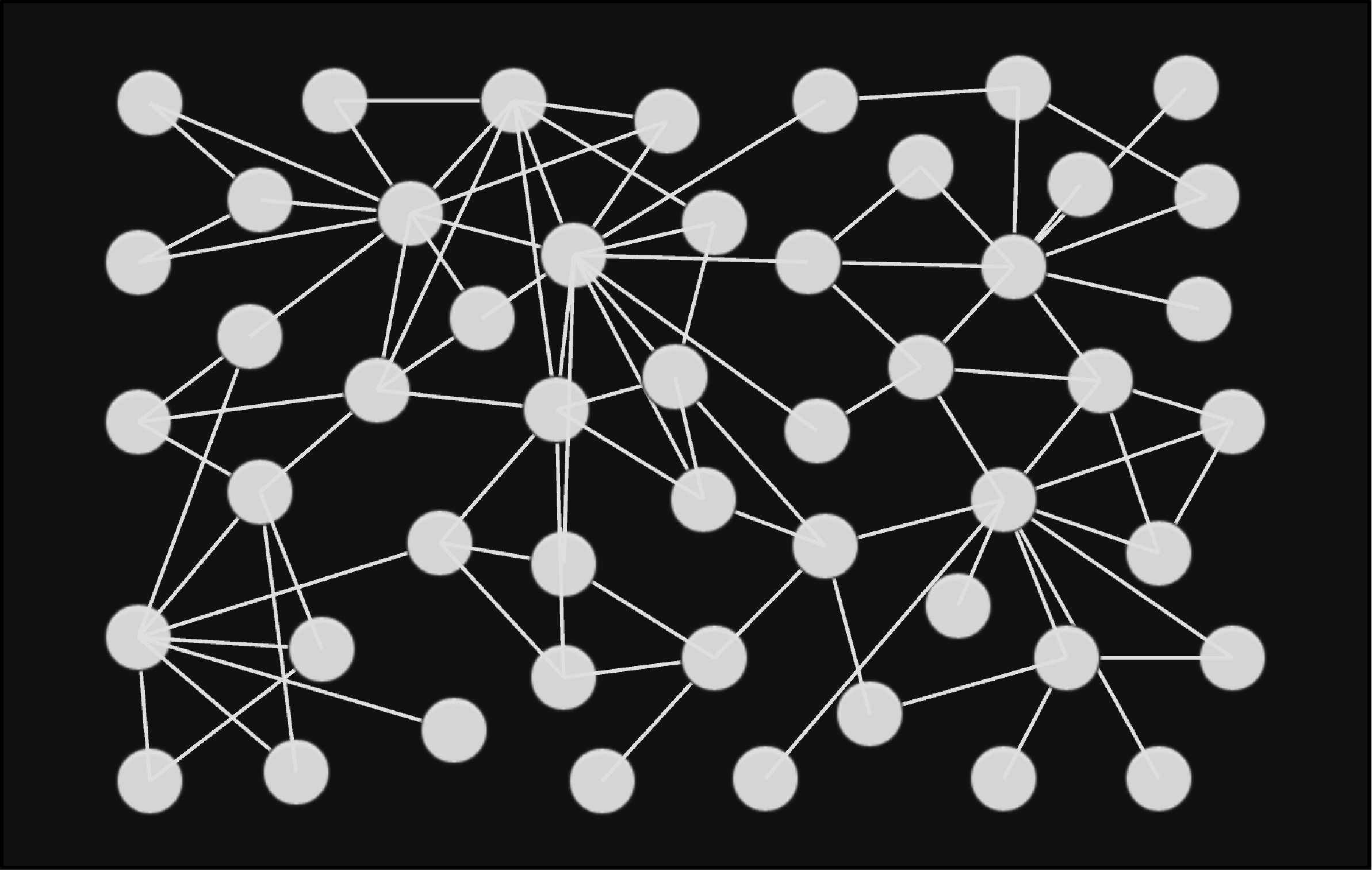} 
        \subcaption{A diagram with maximum complexity.}
        \label{fig:maxdiagram}
    \end{subfigure}
    \hspace{0.003\textwidth}
    \begin{subfigure}[b]{0.23\textwidth}
        \centering
        \includegraphics[width=1\textwidth]{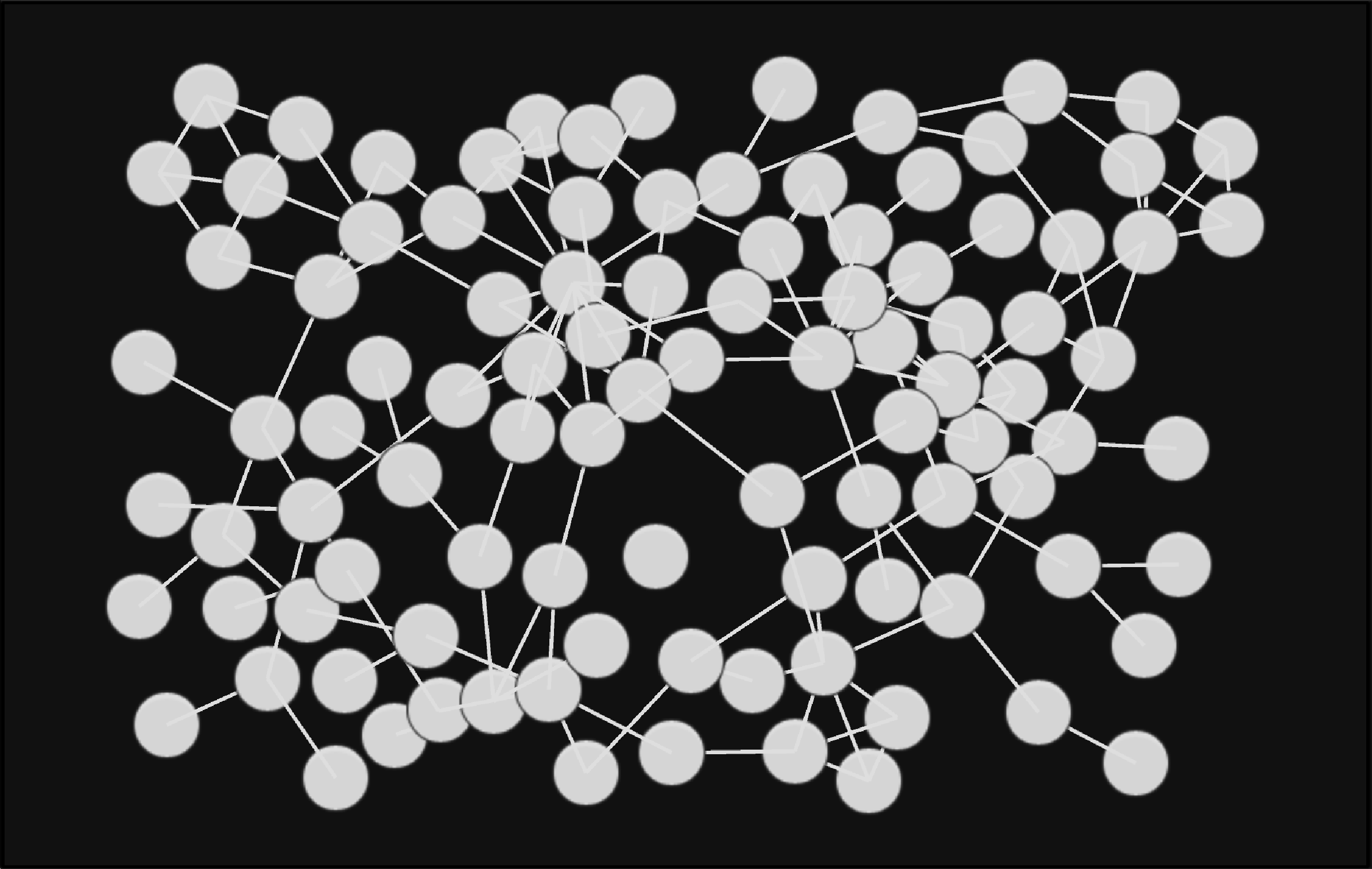}
        \subcaption{A diagram with too much complexity.} 
        \label{fig:impossiblediagram}
    \end{subfigure}

    \caption{Complex diagram examples for TADA.}
    \label{fig:complex_diagram_examples}

    \Description{Here we show two diagrams of high complexity. Sub-figure a) shows the maximum complexity TADA could possibly support, with each node being in proximity to another. This still allows just enough space for the interactions supported by TADA to take place. However, in b) the nodes and links are much more clustered together and even overlapped. This level of complexity makes it impossible to perform the interactions in TADA.}
\end{figure}

Some of the design choices that we made, such as not implementing different zoom levels or having a pannable canvas (e.g.,~\cite{Zeng_Weber_2012,Schmitz_Ertl_2012,Nair_Zhu_Smith_2023}) constrain the complexity of diagrams displayable by TADA. We chose not to support panning or zooming because we think that they would prevent users from constructing a useful spatial mental map of the content [\textbf{DP1}] (this also applies to maps~\cite{Ducasse_Brock_Jouffrais_2018}, which our formative study participants' experiences also confirmed). Nevertheless, it is realistic to expect more complex diagrams (e.g., Figure~\ref{fig:maxdiagram}b) which would be difficult to explore from small screen spaces. Figure~\ref{fig:maxdiagram} shows the most complex diagram that we have found, informally, to still be interactable with in TADA with a 10.6" tablet. In the future, research is required to better understand how spatial understanding can be retained while enabling the exploration of larger pieces of information, especially on smaller displays like tablets and mobile phones.

\subsection{Applicability of TADA} 

The current version of TADA loads diagrams described in GraphML \cite{GraphML_2010}. Although most existing diagrams are not in this format, we expect the translation from other formats, including raster images, to be addressed by advances in computer vision (see, e.g.,~\cite{Kembhavi_Salvato_Kolve_Seo_Hajishirzi_Farhadi_2016, Kim_Yoo_Kim_Lee_Kwak_2018, OpenAI_2023}). TADA is currently designed to represent node-link diagrams. We believe that most diagrams can be represented in the graph structures offered by node-link diagrams. For example, a genealogical tree is a graph with people as nodes and a few types of edges between them. Some other chart types, such as Gantt charts, can also be expressed this way but their representation might be more awkward. 

However, there are three diagram features that TADA does not currently implement and would extend the applicability of the system. First, TADA only supports non-directional links. Second, in its current version, TADA only uses straight links, which can make it inelegant to display some types of layouts. Third, a mechanism to convey the area or region where nodes exist can further facilitate representation through containment.

\section{Conclusion}

We presented insights into approaches for improving diagram accessibility for BLV people. We conducted a formative study and learned about the challenges with diagram access. We summarized these challenges and some workarounds participants employed as steps in a ``ladder of diagram access" that ranges from not knowing that a diagram even exists to taking advantage of a diagram as a representation superior to text.
Informed by our study, we designed TADA, a tool that provides multi-faceted access to node-link diagrams through the careful design and selection of a range of complementary interaction techniques, modes and sound designs that work on an inexpensive tablet. Based on an evaluation study we contribute an understanding of how audio and touch-based interactions can be leveraged to make diagrams more accessible. 



\begin{acks}
We thank Alex Thomo for his invaluable guidance at the initial stages of this work. This research is supported by the University of Victoria, NSERC DG 2020-04401, and the NSERC CREATE VADA graduate program. 
\end{acks}

\bibliographystyle{ACM-Reference-Format}
\bibliography{TADAbib}

\appendix


\section{Formative Study: Participant Details}
\label{appendix:formativeStudy:ParticipantDetails}

Table \ref{tab:interview_participants} presents a summary of participant details in the formative study. 


\begin{table*}[htp]

\caption{Details about participants from the formative study.} 
\label{tab:interview_participants}

\begin{tabular}{ |c|c|c|p{3.5cm}|p{4.5cm}|p{5cm}|}

\hline

\thead{PID}
& \thead{Age} 
& \thead{Gender} 
& \thead{Degree of Vision Loss}
& \thead{Onset}
& \thead{Background}
\\ \hline

P1 
&
35-40 
&
Female 
&
Total 
&
Birth 
&
Employee resource group lead 
\\ \hline

P2 
&
38
&
Male 
&
Total 
&
Young adult 
&
Government employee; Salesperson; Worked in the sports and recreation industry 
\\ \hline

P3 
&
35-40 
&
Female 
&
Total 
&
Birth 
&
Retail 
\\ \hline
 
P4 
&
18 
&
Male 
&
Total 
&
Birth 
&
Student 
\\ \hline

P5 
&
52 
&
Male 
&
Blurry, glasses do not help 
&
Recent years (3 years ago) 
&
Hardware support, IT 
\\ \hline

P6 
&
32 
&
Female 
&
Total 
&
Birth 
&
Government employee 
\\ \hline

P7 
&
60 
&
Male 
&
3\% vision 
&
3 years old 
&
Teacher 
\\ \hline

P8 
&
60 
&
Male 
&
Total 
&
7 years old 
&
Businessman 
\\ \hline

P9 
&
40 
&
Female 
&
Congenital Rubella Syndrome and Glaucoma. Fluctuating vision 
&
Lifelong progression 
&
Human resources, volunteer resources 
\\ \hline

P10 
&
63 
&
Male 
&
1\% vision 
&
Birth 
&
Banking and assistive technology 
\\ \hline

P11 
&
69 
&
Male 
&
Total 
&
Early seventies  
&
Self-employed 
\\ \hline

P12 
&
56 
&
Female 
&
Legally blind, low vision 
&
Losing vision for the past decade  
&
Professor 
\\ \hline

P13 
&
62 
&
Female 
&
Total 
&
2 years old 
&
Information officer, now retired 
\\ \hline

P14 
&
42 
&
Female 
&
Legally blind, significant vision loss 
&
Birth, lifelong progression 
&
Research and teaching 
\\ \hline

P15 
&
50 
&
Male 
&
Total 
&
Vision loss from birth, total vision loss from 2 years ago 
&
Wood products engineer 
\\ \hline 
\end{tabular}

\end{table*}



\section{Formative Study: Interview Questions}
\label{appendix:formativeStudy:InterviewDetails}


Here is the list of interview questions asked for each scenario of a diagram encounter: 

\begin{enumerate}
    
\item[I.] Describe briefly the diagrammatic information from your encounter.
\item[II.] Describe briefly the context of the encounter in terms of when, where, how, who, and
why.
\item[III.] To which degree were you successful in accessing the diagrammatic information? (from 0, not successful at all, to 5, perfectly successful)
\item[IV.] How often do you encounter such a scenario?
\item[V.] What were the motives for or objectives of accessing the information?
\item[VI.] What was the importance of the information?
\item[VII.] Was it an individual, collaborative, or hybrid setting?
\item[VIII.] What was the surrounding physical environment? For example, was it public or
private? Was it personal or professional?
\item[IX.] What media was the information presented in? Was it printed, electronic, verbal, or
other?
\item[X.] What did you do when you accessed the information? What tools did you use to access
the information, if any?
\item[XI.] What types of tasks do you do with the diagrammatic information? (Some examples include getting a general idea, searching, exploring details, browsing, navigating, and annotating. )
\item[XII.] What went well and smoothly when trying to access the diagram?
\item[XIII.] What were the main challenges that you encountered when trying to access the diagram?
\item[XIV.] How did you try to address these challenges, if at all?
\item[XV.] Do you have any ideas that would help you access diagrammatic information in better
ways?

\end{enumerate}


\section{TADA's Implementation Details}
\label{appendix:implementationDetails}


TADA is built using Unity ~\cite{Unity_Technologies}, ChucK~\cite{Wang_Cook_2003}, and Chunity \cite{Atherton_Wang}. Unity, a game engine, uses C\# scripts to interpret diagram data, construct the user interface and interactive diagram components, identify various gestures and their interactions with diagrams, and initiate the corresponding audio feedback. Unity’s multi-platform support allows the system to operate on tablets with different operating systems \cite{Unity_Technologies_multiplatform}. ChucK, a programming language that generates concurrent, strongly-timed procedural audio, renders TADA’s audio. Chunity, which merges ChucK and Unity, serves as the bridge between the two. Native speech recognition is enabled by open-source codes \cite{j1mmyto9/speech-and-text-unity-ios-android}, allowing speech engines to run on either iOS / iPadOS or Android.

\section{Evaluative Study: Participant Details}
\label{appendix:evalStudy:ParticipantDetails}

Table \ref{tab:evaluation_participants} shows a summary of participant details from the evaluative study.


\begin{table*}[htp]

\caption{Details about participants from the evaluation study.}
\label{tab:evaluation_participants}

\begin{tabular}{ |c|c|p{1cm}|p{3.4cm}|p{3.27cm}|p{2.7cm}|p{3.75cm}| }
\hline

\textbf{PID}
& \textbf{Age}
& \textbf{Gender}
& \thead{Degree of Vision Loss}
& \thead{Onset}
& \textbf{Able to Perceive Di- \hphantom agrams Visually?}
& \thead{Background}
\\ \hline

P1  & 52  & Female& Legally blind, less than 3\% peripheral vision left & 10 years old &No& Massage therapist
\\ \hline
P2  & 48  & Male  & Light perception in left eye, blurred vision in right eye & 2022 &No& Construction business
\\ \hline
P3  & 25  & Male  & Total & Birth &No& Technology-related
\\ \hline
P4  & 56  & Female& Total & 21 years old &No& Senior management
\\ \hline
P5  & 46  & Trans-gender female& Total& Birth&No& Volunteer for non-profit organizations and local communities
\\ \hline
P6  & 53  & Female& Total& Birth&No& Language teacher
\\ \hline
P7  & 70  & Female& Total& 5 years old&No& Program coordinator; physio-therapist
\\ \hline
P8  & 101 & Female& Legally blind, usable vision remaining in left eye. &Gradually losing& Yes& Therapeutic dietitian
\\ \hline
P9  & 73  & Male  & Total& 11 years old&No& IT technologist
\\ \hline
P10 & 52  & Female& Legally blind, RP, no central vision, some peripheral vision left&18 years old&No& Manager
\\ \hline
P11 & 65  & Male  & Total& Birth, light and colour perception until age seven&No& Worker at a disability organization
\\ \hline
P12 & 58  & Female& Legally blind, partial optic atrophy with peripheral vision& Prematurely&Yes (able to visually perceive some)& Unemployed
\\ \hline
P13 & 35  & Female& Legally blind& Birth&Yes (able to visually perceive some)& Program coordinator
\\ \hline
P14 &41 & Male & Light perception in left eye, total blind in right eye & Birth &No& Assistive tech support
\\ \hline
P15 &56 & Female& Legally blind, 3\% vision in right eye. & Birth, gradual vision loss&Yes (able to visually perceive some)& Previously professional musician; After vision loss, teaching music and assistive technology
\\ \hline
P16 &46 & Female& Legally blind, totally blind in right eye, some narrow vision in left eye & Birth &No& Customer service and sales
\\ \hline
P17 &32 & Female& Total & Birth &No& Government work
\\ \hline
P18 &50 & Female& Legally blind &2013 &No& Medical assistant
\\ \hline
P19 &58 & Male & RP, legally blind &2008 &Yes& Accounting
\\ \hline
P20 &37 & Female& Legally blind&9 years old&No& Non-profit sector
\\ \hline
P21 &68 & Male& Legally blind, some peripheral vision left&2014&No& Previously transit driver
\\ \hline
P22 &61 & Male& Legally blind, some peripheral vision left&30 years old&No& Software engineering
\\ \hline
P23 &38 & Male& Total& Young adult&No& Government employee; Salesperson; Worked in the sports and recreation industry
\\ \hline
P24 &28 & Male& Legally blind&2016&No& Assistive tech support
\\ \hline
P25 &49 & Female& Legally blind, can only see light and shadow&Birth, gradual vision loss&No& Executive director at non-profit organization
\\ \hline
\end{tabular}
\end{table*}



\section{Evaluative Study: Experimental Blocks}
\label{appendix:evalStudy:TaskDetails}

This appendix presents the details of the experimental blocks from the evaluative study. 

The 6 experimental blocks each focused on an interaction technique. The interaction techniques evaluated are (1) Single-finger Sweep; (2) Five-finger Dome; (3) Dwell+Tap; (4) Single-finger Flick-down, Speech Command for Searching and Single-finger Follow; (5) Dwell+Circle; and (6) Dwell+Radiate. We tasked the participants to only use the dedicated interaction technique to answer its associated question(s) and the questions asked about the test diagrams are listed below corresponding to the blocks.

\subsection{Gain an Overview with Single-finger Sweep}

We used test diagram 1 (Figure \ref{fig:device_diag_1}) which was a more complex diagram than the tactile diagram, and it has distinct characteristics in the diagrammatic structure in its four quadrants meant for comparison. We tasked participants to gain an overview with Single-finger Sweep and answer the following questions about the test diagram:
\begin{enumerate}[leftmargin=1.5cm]
    \item [Q1.1:] Describe the overall layout of the diagram.  
    \item [Q1.2:] Which quadrant of the diagram has the most connections between people? (Answer: Bottom left.)
    \item [Q1.3:] Which quadrant has the least connections? (Answer: Top left.)
    \item [Q1.4:] Which quadrant of the diagram has the most people? (Answer: Bottom right.)
    \item [Q1.5:] Which quadrant has the least people? (Answer: Top left.)
\end{enumerate}

For the first question (Q1.1), we asked participants to sweep and explore the entire diagram and provide answers regarding the general layout. We then asked more specific questions (Q1.1 to 1.4) for participants to explicitly think in terms of quadrants and in terms of the connectivity and density of nodes. We asked participants to not count exactly but to gather a general overview. 

\subsection{Gain a Summary of a Sub-region with Five-finger Dome}

To learn about the effectiveness of the Five-finger Dome within sub-regions, we tasked participants to gather the summaries of sub-regions using Five-finger Dome and answer the following questions: 

\begin{enumerate}[leftmargin=1.5cm]
    \item [Q2.1:] Which quadrant of the diagram has the most connections between people? (Answer: Bottom right.)
    \item [Q2.2:] Which quadrant has the least connections? (Answer: Bottom left.)
    \item [Q2.3:] Which quadrant of the diagram has the most people? (Answer: Top right.)
    \item [Q2.4:] Which quadrant has the least people? (Answer: Bottom left.)
    \item [Q2.5:] Which quadrant has people with unequal numbers of friends? (Answer: Top right.) 
\end{enumerate}

We asked the same specific questions about connectivity and density of nodes but switched the diagram to test diagram 2 (Figure \ref{fig:device_diag_2}) to make sure participants cannot reuse the same answers from test diagram 1. It included modifications to test diagram 1 such as changing the positions and numbers of the nodes in each quadrant. Additionally, because we also want to evaluate explicitly the effectiveness of the pacing of audio, Q2.5 was asked to see if participants could pick up the variations in the pacing of the link audio sounds within an area. There can be multiple answers to Q2.5, and we asked participants to identify the most obvious one where there is one person very popular in the top right quadrant while the others are not so that this node's link sounds are paced much faster. Again, we asked participants to not count exactly but to gather a general overview, and we did not evaluate what details this interaction could potentially achieve.

\subsection{Gain Textual Details with Dwell+Tap}

We asked a question about the details of a diagram element from performing the Dwell+Tap interaction: 

\begin{enumerate}[leftmargin=1.5cm]
    \item [Q3:] Given a person, what is the name and professional background associated with it?  
\end{enumerate}

We used test diagram 3 (Figure \ref{fig:device_diag_3}) for this task and the remaining three tasks (Task 4 to Task 6). We do not have to design different diagrams for the remaining interactions because the answers would not affect answering later questions. In Q3, we provided participants with a node to examine that had a name other than ``Bob" which was going to be used in later questions.

\subsection{Search with Single-finger Flick-down, Speech Command and Single-finger Follow}

We tasked the participants to locate a piece of information and answer the following question:

\begin{enumerate}[leftmargin=1.5cm]
    \item [Q4:] Where is Bob? Locate Bob. 
\end{enumerate}

To eliminate other factors affecting the accuracy of speech recognition such as participants' accents and background noise, we conducted a Wizard of Oz experiment where we asked participants to only look for where Bob is in the diagram, and the system is only configured to search for Bob.

\subsection{Explore Connected Links with Dwell+Circle}

We asked participants to perform Dwell+Circle to identify the number of connected links of a node: 

\begin{enumerate}[leftmargin=1.5cm]
    \item [Q5:] How many friends does Bob have? (Answer: 4.)
\end{enumerate}

Dwell+Circle aims to explore a node's connected links in detail, gathering the number of connections, their directions and their relative distances. Q5 only asked about the number of connections because, in the interest of time, we decided to focus on whether one could complete a full circle and count the number of connected links without exploring other details. Q6 below partially explored the directional information of the connected links.

\subsection{Navigate with Dwell+Radiate}

Following Task 5, we asked participants to navigate to any nodes of the neighbouring node using Dwell+Radiate:

\begin{enumerate}[leftmargin=1.5cm]
    \item [Q6:] Navigate from Bob to any of his friends.  
\end{enumerate}

Dwell+Radiate works in combination with Dwell+Circle. Using the Dwell+Circle interaction to explore the connected links, one can locate a link of interest and then decide to navigate toward its direction using the Dwell+Radiate interaction. Therefore, Q6 evaluated the Dwell+Circle interaction partially and the Radiate interaction.



\end{document}